\documentclass[journal, onecolumn]{IEEEtran}

\usepackage{graphicx}          
\usepackage{amsmath} 
\usepackage[english]{babel}
\usepackage{amsfonts}
\usepackage{amstext}
\usepackage{caption}
\usepackage{subcaption}
\usepackage{tabularx}
\usepackage{savesym}
\savesymbol{AND}
\usepackage{algorithm}
\usepackage{algorithmic}
\usepackage{color}

\usepackage{exStyle}

\newtheorem{theorem}{Theorem}[section]
\newtheorem{lemma}[theorem]{Lemma}
\newtheorem{proposition}[theorem]{Proposition} 

\newtheorem{assumption}[theorem]{Assumption}
\newtheorem{remark}[theorem]{Remark}
\newtheorem{problem}[theorem]{Problem}

\begin{document}
%

\title{Quantization Design for Distributed Optimization}
%
%
%


\author{Ye Pu, Melanie N. Zeilinger and Colin N. Jones%
\thanks{Y. Pu and C.N. Jones are with the Automatic Control Lab, \'Ecole Polytechnique F\'ed\'erale de Lausanne, EPFL-STI-IGM-LA
Station 9 CH-1015 Lausanne, Switzerland, e-mail: {\tt \small \{y.pu,colin.jones\}@epfl.ch}. }%
\thanks{M.N. Zeilinger is with the Empirical Inference Department, Max Planck Institute for Intelligent Systems, 72076 T\"ubingen, Germany, e-mail: {\tt\small melanie.zeilinger@tuebingen.mpg.de}. }%
\thanks{This work has received funding from the European Research Council under the European Union's Seventh Framework Programme (FP/2007-2013)/ ERC Grant Agreement n. 307608. The research of M. N. Zeilinger has received funding from the EU FP7 under grant agreement no. PIOF-GA-2011-301436-``COGENT''.}%
}




\maketitle

\begin{abstract}
We consider the problem of solving a distributed optimization problem using a distributed computing platform, where the communication in the network is limited: each node can only communicate with its neighbours and the channel has a limited data-rate. A common technique to address the latter limitation is to apply quantization to the exchanged information. We propose two distributed optimization algorithms with an iteratively refining quantization design based on the inexact proximal gradient method and its accelerated variant. We show that if the parameters of the quantizers, i.e. the number of bits and the initial quantization intervals, satisfy certain conditions, then the quantization error is bounded by a linearly decreasing function and the convergence of the distributed algorithms is guaranteed. Furthermore, we prove that after imposing the quantization scheme, the distributed algorithms still exhibit a linear convergence rate, and show complexity upper-bounds on the number of iterations to achieve a given accuracy. Finally, we demonstrate the performance of the proposed algorithms and the theoretical findings for solving a distributed optimal control problem. 
\end{abstract}

%
\IEEEpeerreviewmaketitle

\section{Introduction}\label{sec:Introduction}

 Distributed optimization methods for networked systems that have many coupled sub-systems and must act based on local information, are critical in many engineering problems, e.g. resource allocation, distributed estimation and distributed control problems. The algorithms are required to solve a global optimization problem in a distributed fashion subject to communication constraints. 

 Inexact distributed optimization methods are attracting increasing attention, since these techniques have the potential to deal with errors, for instance caused by inexact solution of local problems as well as unreliable or limited communication, e.g., transmission failures and quantization errors. Previous work has aimed at addressing the questions of how such errors affect the algorithm and under what conditions the convergence of the distributed algorithms can be guaranteed. In \cite{dinh_fast_2012}, the authors propose an inexact decomposition algorithm for solving distributed optimization problems by employing smoothing techniques and an excessive gap condition. In our previous work \cite{pu_inexact_2014}, we have proposed an inexact splitting method, named the inexact fast alternating minimization algorithm, and applied it to distributed optimization problems, where local computation errors as well as errors resulting from limited communication are allowed, and convergence conditions on the errors are derived based on a complexity upper-bound. Some other related references for inexact optimization algorithms include \cite{devolder_first_order_2013}, \cite{nedelcu_complexity_2014} and \cite{schmidt_convergence_2011}. In \cite{schmidt_convergence_2011}, an inexact proximal-gradient method, as well as its accelerated version, are introduced. The proximal gradient method, also known as the iterative shrinkage-thresholding algorithm (ISTA) \cite{beck_fISTA_2009}, has two main steps: the first one is to compute the gradient of the smooth objective and the second one is to solve the proximal minimization. The conceptual idea of the inexact proximal-gradient method is to allow errors in these two steps, i.e. an error in the calculation of the gradient and an error in the proximal minimization. The results in \cite{schmidt_convergence_2011} show convergence properties of the inexact proximal-gradient method and provide conditions on the errors, under which convergence of the algorithm can be guaranteed. 

 We consider a distributed optimization problem, where each sub-problem has a local cost function that involves both local and neighbouring variables, and local constraints on local variables. The problem is solved in a distributed manner with only local communication, i.e. between neighbouring sub-systems. In addition, the communication bandwidth between neighbouring sub-systems is limited. In order to meet the limited communication data-rate, the information exchanged between the neighbouring sub-systems needs to be quantized. The quantization process results in inexact iterations throughout the distributed optimization algorithm, which effects its convergence. Related work includes \cite{carli_average_2007}, \cite{akshay_quantized_2007}, \cite{thanou_distributed_2013} and \cite{nedic_distributed_2008}, which study the effects of quantization on the performance of averaging or distributed optimization algorithms.

 We propose two distributed optimization algorithms with progressive quantization design building on the work in \cite{schmidt_convergence_2011} and \cite{thanou_distributed_2013}. The main idea behind the proposed methods is to apply the inexact gradient method to the distributed optimization problem and to employ the error conditions, which guarantee convergence to the global optimum, to design a progressive quantizer. Motivated by the linear convergence upper-bound of the optimization algorithm, the range of the quantizer is set to reduce linearly at a rate smaller than one and larger than the rate of the algorithm, in order to refine the information exchanged in the network with each iteration and achieve overall converge to the global optimum. The proposed quantization method is computationally cheap and consistent throughout the iterations as every node implements the same quantization procedure. 

 This work extends the initial ideas presented in \cite{pu_quantization_2015} for designing a quantization scheme for unconstrained distributed optimization. In particular, the paper makes the following main extensions and contributions:  

\begin{itemize}
\item Constrained optimization problems: We consider distributed optimization problems with convex local constraints. To handle the constraints, two projection steps are required. One is applied before the information exchange, and the other after. The reason to have a second projection is that after the information exchange, the quantized value received by each agent can be an infeasible solution subject to the local constraints. The second projection step therefore guarantees that at each iteration every agent has a feasible solution for the computation of the gradient. We present conditions on the number of bits and the initial quantization intervals, which guarantee convergence of the algorithms. We show that after imposing the quantization scheme including the two projections, the algorithms preserve the linear convergence rate, and furthermore derive complexity upper-bounds on the number of iterations to achieve a given accuracy. In addition, we provide a discussion about how the minimum number of bits and the corresponding minimum initial quantization intervals can be obtained.
\item Accelerated algorithm: We propose an accelerated variant of the distributed optimization algorithm with quantization refinement based on the inexact accelerated proximal-gradient method. With the acceleration step, the algorithm preserves the linear convergence rate, but the constant of the rate will be improved.
\item Distributed optimal control example: We demonstrate the performance of the proposed method and the theoretical results for solving an distributed optimal control example. 
\end{itemize}

\section{Preliminaries}\label{sec:Preliminaries}
\subsection{Notation}
Let $v\in\mathbb{R}^{n_v}$ be a vector. $\|v\|$ and $\|v\|_{\infty}$ denote the $l_2$ and infinity norms of $v$, respectively. Note that $\|v\|_{\infty}\leq \|v\|_{2}\leq \sqrt{n_v}\|v\|_{\infty}$. Let $\mathbb{C}$ be a subset of $\mathbb{R}^{n_v}$. The projection of any point $v\in\mathbb{R}^{n_v}$ onto the set $\mathbb{C}$ is denoted by $\mbox{Proj}_{\mathbb{C}}(v):=\mbox{argmin}_{\mu\in\mathbb{C}}\;\|\mu-v\|$. Let $f: \Theta \rightarrow \Omega$ be a strongly convex function; $\sigma_f$ denotes the convexity modulus $f(v)\geq f(\mu) + \left< \partial f(\mu), v-\mu\right> + \frac{\sigma_f}{2} \|v-\mu\|^{2}$ for any $v,\mu\in \Theta$, where $\partial f(\cdot)$ denotes the set of sub-gradients of the function $f$ at a given point. $L(f)$ denotes a Lipschitz constant of the function $f$, i.e. $\|f(v)-f(\mu)\|\leq L(f)\|v - \mu \|$, $\forall v,\mu\in \Theta$. The proximity operator is defined as
\begin{equation}\label{eq:proximity operator}
\mbox{prox}_{f}(v) = \mbox{argmin}_{w} \quad f(w) + \frac{1}{2}\|w-v\|^{2}\enspace .
\end{equation}
We refer to \cite{Bertsekas_Convex_2003} and \cite{horn_matrix_1990} for details on the definitions and properties above. The proximity operator with an extra subscript $\epsilon$, i.e. $\mu=\mbox{prox}_{f,\epsilon}(v)$, means that a maximum computation error $\epsilon$ is allowed in the proximal objective function:
\begin{equation}\label{eq:epsilon error in proximity operator}
f(\mu) + \frac{1}{2}\|\mu-v\|^{2} \; \leq\; \epsilon + \mbox{min}_{w} \left\lbrace f(w) + \frac{1}{2}\|w-v\|^{2}\right\rbrace
\end{equation}

\subsection{Inexact Proximal-Gradient Method}\label{se:inexact PGM}
In this section, we will introduce the inexact proximal-gradient method (inexact PGM) proposed in \cite{schmidt_convergence_2011}. Inexact PGM is presented in Algorithm~\ref{al:inexact ISTA}. It addresses optimization problems of the form given in Problem~\ref{pr:problem ISTA} and requires Assumption~\ref{as:ISTA} for convergence with a linear rate. 
\begin{problem}\label{pr:problem ISTA}
\begin{align*}
\min_{x\in \mathbb R^{n_x}} \quad \Phi(x) =\phi(x)+\psi(x)\enspace .
\end{align*}
\end{problem}
\begin{assumption}\label{as:ISTA}
\begin{itemize}
\item $\phi$ is a strongly convex function with a convexity modulus $\sigma_{\phi}$ and Lipschitz continuous gradient with Lipschitz constant $L(\nabla \phi)$.
\item $\psi$ is a lower semi-continuous convex function, not necessarily smooth.
\end{itemize}
\end{assumption}
\begin{scriptsize}
\begin{algorithm}
\caption{Inexact Proximal-Gradient Method}
\begin{algorithmic}
\REQUIRE Require $x^{0} \in \mathbb R^{n_x}$ and $\tau<\frac{1}{L(\nabla \phi)}$
\FOR {$k=0,1,\cdots$}
\STATE $x^{k+1} =  \mbox{prox}_{\tau \psi, \epsilon^{k}}(x^{k} - \tau (\nabla \phi(x^{k}) + e^k))$
\ENDFOR
\end{algorithmic}
\label{al:inexact ISTA}
\end{algorithm}\begin{footnotesize}
\end{footnotesize}
\end{scriptsize}

Inexact PGM in Algorithm~\ref{al:inexact ISTA} allows two kinds of errors: $\{e^{k}\}$ represents the error in the gradient calculations of $\phi$, and $\{\epsilon^{k}\}$ represents the error in the computation of the proximal minimization in (\ref{eq:epsilon error in proximity operator}) at every iteration $k$. The following proposition states the convergence property of inexact PGM.
%
%
\begin{proposition}[Proposition 3 in \cite{schmidt_convergence_2011}]\label{pr:convergence rate of inexact ISTA}
Let $\{x^k\}$ be generated by inexact PGM defined in Algorithm~\ref{al:inexact ISTA}. If Assumption \ref{as:ISTA} holds, then for any $k\geq 0$ we have:
\begin{align}
\|x^{k+1}-x^{\star}\|\leq (1-\gamma)^{k+1}\cdot(\|x^{0}-x^{\star}\|+\Gamma^{k})\enspace ,
\end{align}
where $\gamma = \frac{\sigma_{\phi}}{L(\nabla \phi)}$ and $x^{0}$ and $x^{\star}$ denote the initial sequence of  Algorithm~\ref{al:inexact ISTA} and the optimal solution of Problem~\ref{pr:problem ISTA}, respectively, and 
\begin{equation*}
\Gamma^{k} = \sum^{k}_{p=0} (1-\gamma)^{-p-1}\cdot\left(\frac{1}{L(\nabla \phi)} \|e^{p}\| +\sqrt{\frac{2}{L(\nabla \phi)}}\sqrt{\epsilon^{p}}\right)\enspace .
\end{equation*} 
\end{proposition}

 As discussed in \cite{schmidt_convergence_2011}, the upper-bound in Proposition~\ref{pr:convergence rate of inexact ISTA} allows one to derive sufficient conditions on the error sequences $\{e^{k}\}$ and $\{\epsilon^{k}\}$ for convergence of the algorithm to the optimal solution $x^*$, where $\mu = 1-\gamma$:
\begin{itemize}\label{re:the decreasing rate of the errors in inexact PGM}
\item If the series $\{\|e^{k}\|\}$ and $\{\sqrt{\epsilon^{k}}\}$ decrease at a linear rate with the constant $\kappa < \mu$, then $\|x^{k}-x^{\star}\|$ converges at a linear rate with the constant $\mu$.
\item If the series $\{\|e^{k}\|\}$ and $\{\sqrt{\epsilon^{k}}\}$ decrease at a linear rate with the constant $\mu < \kappa <1$, then $\|x^{k}-x^{\star}\|$ converges at the same rate with the constant $\kappa$. 
\item If the series $\{\|e^{k}\|\}$ and $\{\sqrt{\epsilon^{k}}\}$ decrease at a linear rate with the constant $\kappa = \mu$, then $\|x^{k}-x^{\star}\|$ converges at a rate of $O(k\cdot\mu^{k})$. 
\end{itemize}
\begin{remark}
Compared to \cite{schmidt_convergence_2011}, we modify the index of the sequence in Algorithm~\ref{al:inexact ISTA} from $x_k$ to $x_{k+1}$ and the corresponding index in Proposition~\ref{pr:convergence rate of inexact ISTA}, such that in Section~\ref{se:Distributed optimization with limited communication} the quantization errors have the same index as the quantized sequences.  
\end{remark}
\subsection{Inexact Accelerated Proximal-Gradient Method}\label{se:inexact APGM}
 In this section, we introduce an accelerated variant of inexact PGM, named the inexact accelerated proximal-gradient method (inexact APGM) proposed in \cite{schmidt_convergence_2011}. Compared to inexact PGM, it addresses the same problem class in Problem~\ref{pr:problem ISTA}  and requires the same assumption in Assumption~\ref{as:ISTA} for linear convergence, but involves one extra linear update in Algorithm 2, which improves the constant of the linear convergence rate from $(1-\gamma)$ to $\sqrt{1-\sqrt{\gamma}}$.
\begin{scriptsize}
\begin{algorithm}
\caption{Inexact Accelerated Proximal-Gradient Method}
\begin{algorithmic}
\REQUIRE Initialize $x^{0} = y^{0} \in \mathbb R^{n_x}$ and $\tau<\frac{1}{L(\nabla \phi)}$
\FOR {$k=0,1,\cdots$}
\STATE $x^{k+1} =  \mbox{prox}_{\tau \psi, \epsilon^{k}}(y^{k} - \tau (\nabla \phi(y^{k}) + e^k))$
\STATE $y^{k+1} = x^{k+1} + \frac{1-\sqrt{\gamma}}{1+\sqrt{\gamma}}(x^{k+1}-x^{k})$
\ENDFOR
\end{algorithmic}
\label{al:inexact APGM}
\end{algorithm}\begin{footnotesize}
\end{footnotesize}
\end{scriptsize}

Proposition~4 of \cite{schmidt_convergence_2011} presents a complexity upper-bound on the sequence $\{\Phi (x^{k+1}) - \Phi (x^{\star})\}$, where the sequence $\{x^{k+1}\}$ is generated by inexact APGM. The following proposition extends this result and states a complexity upper-bound on $\|x^{k+1}-x^{\star}\|$.

\begin{proposition}\label{pr:convergence rate of inexact APGM}
Let $\{x^k\}$ be generated by inexact APGM defined in Algorithm~\ref{al:inexact APGM}. If Assumption \ref{as:ISTA} holds, then for any $k\geq 0$ we have:
\begin{equation}\label{eq:complexity bound for inexact APGM}
\|x^{k+1}-x^{\star}\|\leq (1-\sqrt{\gamma})^{\frac{k+1}{2}}\cdot\left(\frac{2\sqrt{\Phi(x^{0})-\Phi(x^{\star})}}{\sqrt{\sigma_{\phi}}}+\Theta^{k}\right)\enspace ,
\end{equation}
where $\gamma = \frac{\sigma_{\phi}}{L(\nabla \phi)}$, $x^{0}$ and $x^{\star}$ denote the initial sequence of  Algorithm~\ref{al:inexact ISTA} and the optimal solution of Problem~\ref{pr:problem ISTA}, respectively, and 
\begin{align*}
\Theta^{k} = \frac{2}{\sigma_{\phi}}\cdot\sum^{k}_{p=0} (1-\sqrt{\gamma})^{\frac{-p-1}{2}} \cdot\bigg(\|e^{p}\| + (\sqrt{2L(\nabla \phi)}+\sqrt{\frac{\sigma_{\phi}}{2}})\cdot \sqrt{\epsilon^{p}}\bigg)\enspace .
\end{align*} 
\end{proposition}

 The proof of Proposition~\ref{pr:convergence rate of inexact APGM} will be given in the appendix in Section~\ref{SEC: appendix proof of Proposition of convergence rate of inexact APGM}. The upper-bound in Proposition~\ref{pr:convergence rate of inexact APGM} provides similar sufficient conditions on the error sequences $\{e^{k}\}$ and $\{\epsilon^{k}\}$ for the convergence of Algorithm~\ref{al:inexact APGM}, which are obtained by replacing $\mu = 1-\gamma$ in the sufficient conditions for Algorithm~\ref{al:inexact ISTA} in Section~\ref{se:inexact PGM} with $\mu = \sqrt{1-\sqrt{\gamma}}$.

\subsection{Uniform quantizer}\label{se:uniform quantizer}
Let $x$ be a real number. A uniform quantizer with a quantization step-size $\Delta$ and the mid-value $\bar{x}$ can be expressed as
\begin{equation}\label{eq: uniform quantizer}
Q(x) = \bar{x} + \mbox{sgn}(x-\bar{x})\cdot\Delta\cdot \left\lfloor\frac{\|x-\bar{x}\|}{\Delta}+\frac{1}{2}\right\rfloor \enspace , 
\end{equation}
where $\mbox{sgn}(\cdot)$ is the sign function. The parameter $\Delta$ is equal to $\Delta = \frac{l}{2^{n}}$, where $l$ represents the size of the quantization interval and $n$ is the number of bits sent by the quantizer. In this paper, we assume that $n$ is a fixed number, which means that the quantization interval is set to be $[\bar{x}-\frac{l}{2}, \bar{x} + \frac{l}{2}]$. The quantization error is upper-bounded by
\begin{equation}\label{eq:quantization error}
\|x-Q(x)\| \leq \frac{\Delta}{2} = \frac{l}{2^{n+1}} \enspace . 
\end{equation}
For the case that the input of the quantizer and the mid-value are not real numbers, but vectors with the same dimension $n_{x}$, the quantizer $Q$ is composed of $n_x$ independent scalar quantizers in~(\ref{eq: uniform quantizer}) with the same quantization interval $l$ and corresponding mid-value. In this paper, we design a uniform quantizer denoted as $Q^{k}(\cdot)$ with changing quantization interval $l^{k}$ and mid-value $\bar{x}^{k}$ at every iteration $k$ of the optimization algorithm.

\section{Distributed optimization with limited communication}\label{se:Distributed optimization with limited communication}
 In this section, we propose two distributed optimization algorithms with progressive quantization design based on the inexact PGM algorithm and its accelerated variant. The main challenge is that the communication in the distributed optimization algorithms is limited and the information exchanged in the network needs to be quantized. We propose a progressive quantizer with changing parameters, which satisfies the communication limitations, while ensuring that the errors induced by quantization satisfy the conditions for convergence. 

\subsection{Distributed optimization problem}\label{se:distributed optimization problem}

In this paper, we consider a distributed optimization problem on a network of $M$ sub-systems (nodes). The sub-systems communicate according to a fixed undirected graph $G =(\mathcal{V},\mathcal{E})$. The vertex set $\mathcal{V} = \{1,2,\cdots,M\}$ represents the sub-systems and the edge set $\mathcal{E}\subseteq \mathcal{V}\times \mathcal{V}$ specifies pairs of sub-systems that can communicate. If $(i,j)\in \mathcal{E}$, we say that sub-systems $i$ and $j$ are neighbours, and we denote by $\mathcal{N}_i = \{j| (i,j)\in \mathcal{E}\}$ the set of the neighbours of sub-system $i$. Note that $\mathcal{N}_i$ includes $i$. We denote $d$ as the degree of the graph $G$. The optimization variable of sub-system $i$ and the global variable are denoted by $x_i$ and $x=[x^{T}_1,\cdots,x^{T}_M]^{T}$, respectively. For each sub-system $i$, the local variable has a convex local constraint $x_{i}\in \mathbb{C}_{i}\subseteq \mathbb{R}^{n_{m_i}}$. The constraint on the global variable $x$ is denoted by $\mathbb{C}=\prod_{1\leq i\leq M} \mathbb{C}_i$. The dimension of the local variable $x_i$ is denoted by $m_i$ and the maximum dimension of the local variables is denoted by $\bar{m}$, i.e. $\bar{m} := \max_{1\leq i \leq M} m_i$. The concatenation of the variable of sub-system $i$ and the variables of its neighbours is denoted by $x_{\mathcal{N}_i}$, and the corresponding constraint on $x_{\mathcal{N}_i}$ is denoted by $\mathbb{C}_{\mathcal{N}_i}=\prod_{j\in \mathcal{N}_i} \mathbb{C}_j$. With the selection matrices $E_i$ and $F_{ji}$, they can be represented as $x_{\mathcal{N}_{i}} = E_ix$ and $x_i = F_{ji}x_{\mathcal{N}_{j}}$, $j\in\mathcal{N}_i $, which implies the relation between the local variable $x_i$ and the global variable $x$, i.e. $x_i = F_{ji}E_jx$, $j\in\mathcal{N}_i $. Note that $E_i$ and $F_{ji}$ are selection matrices, and therefore $\|E_i\|=\|F_{ji}\|=1$. We solve a distributed optimization problem of the formulation in Problem~\ref{pr:distributed optimization}:
\begin{problem}\label{pr:distributed optimization}
\begin{align*}
\min_{x,\; x_{\mathcal{N}_{i}}} & \quad f(x) =\sum^{M}_{i=1} f_{i}(x_{\mathcal{N}_{i}})\\
s.t. &\quad  x_i\in \mathbb{C}_{i}\enspace , \; x_{i} = F_{ji}x_{\mathcal{N}_{j}}\enspace , \; j\in\mathcal{N}_i \enspace , \; x_{\mathcal{N}_{i}} = E_ix\enspace , \;  i=1,2,\cdots ,M\enspace .
\end{align*}
\end{problem}


\begin{assumption}\label{as: assumption on f}
We assume that the global cost function $f(\cdot)$ is strongly convex with a convexity modulus $\sigma_{f}$ and Lipschitz continuous gradient with Lipschitz constant $L$, i.e. $\|\nabla f(x_1) - \nabla f(x_2)\| \leq L\|x_1 - x_2\|$ for any $x_1$ and $x_2$.
%
%
\end{assumption} 
\begin{assumption}\label{as: Ci is convex, convex local constraints}
The local constraint $\mathbb{C}_{i}$ is a convex set, for $i = 1,\cdots , M$.
\end{assumption}
\begin{assumption}\label{as: assumption on fi}
We assume that every local cost function $f_i(\cdot)$ has Lipschitz continuous gradient with Lipschitz constant $L_i$, and denote $L_{max}$ as the maximum Lipschitz constant of the local functions, i.e. $L_{max} := \max_{1\leq i \leq M} L_i$.
\end{assumption}
%
%

\subsection{Qualitative description of the algorithm}

In this section, we provide a qualitative description of the distributed optimization algorithm with quantization refinement to introduce the main idea of the approach. We apply the inexact PGM algorithm to the distributed optimization problem in Problem~\ref{pr:distributed optimization}, where the two objectives in Problem~\ref{pr:problem ISTA} are chosen as $\phi = \sum^{M}_{i=1} f_{i}(x_{\mathcal{N}_i})$ and $\psi = \sum^{M}_{i=1} I_{\mathbb{C}_i}(x_i)$, where $I_{\mathbb{C}_i}$ denotes the indicator function on the set $\mathbb{C}_i$. The parameter $\gamma$ is equal to
\begin{equation}\label{eq:gamma}
\gamma = \frac{\sigma_f}{L}\enspace .
\end{equation}

\begin{figure}
        \centering
        \begin{subfigure}[b]{0.56\textwidth}
                \centering
                \includegraphics[width=\textwidth]{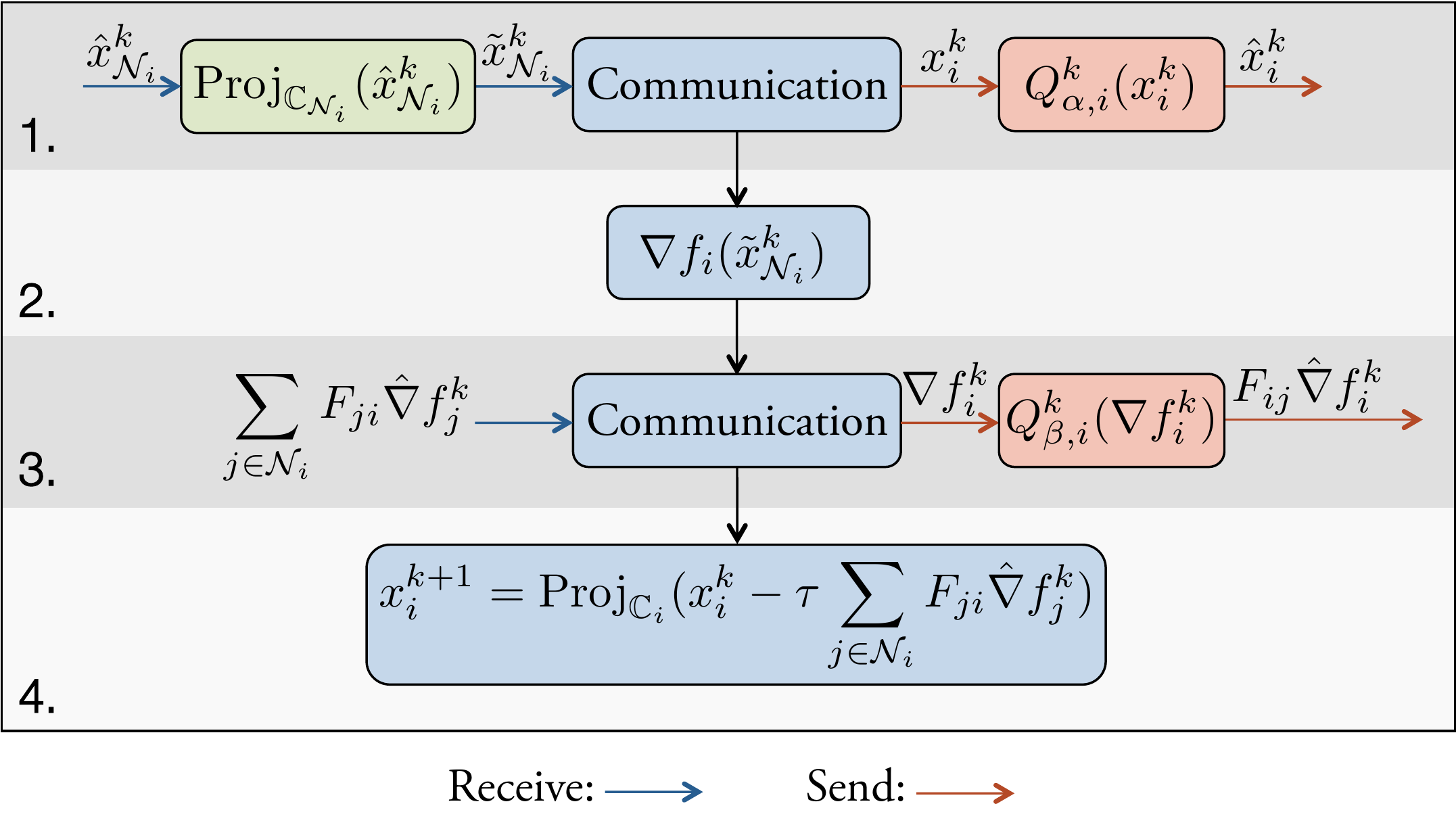}
        \end{subfigure}
        \caption{Distributed algorithm with quantization refinement for subsystem $i$ at iteration $k$}
        \label{fig:distributed optimization with limited bits}
\end{figure}

 The communication in the network is limited: each sub-system in the network can only communicate with its neighbours, and at each iteration, only a fixed number of bits can be transmitted. Only considering the first limitation, the distributed optimization algorithm resulting from applying the inexact PGM algorithm to Problem~\ref{pr:distributed optimization} is represented by the blue boxes in Fig.~\ref{fig:distributed optimization with limited bits}. At iteration $k$, sub-system $i$ carries out four main steps: 
 \begin{enumerate}
 \item[1.]Send the local variable to its neighbours;
 \item[2.]Compute the local gradient;
 \item[3.]Send the local gradient to its neighbours;
 \item[4.]Update the local variable and compute the projection of the updated local variable on the local constraint. 
 \end{enumerate}

 To handle the second limitation, we design two uniform quantizers (the salmon-pink boxes) for the two communication steps for each sub-system $Q^{k}_{\alpha ,i}$ and $Q^{k}_{\beta ,i}$ using a varying quantization interval and mid-value to refine the exchanged information at each iteration. Motivated by the second sufficient condition on the error sequences $\{e^{k}\}$ and $\{\epsilon^{k}\}$ for the convergence of the inexact PGM algorithm discussed in Section~\ref{se:inexact PGM} (if the sequences $\{\|e^{k}\|\}$ and $\{\sqrt{\epsilon^{k}}\}$ decrease at a linear rate with the constant $(1-\gamma)< \kappa <1$, then $\|x^{k}-x^{\star}\|$ converges with the same rate), the quantization intervals are set to be a linearly decreasing function $l^{k}_{\alpha,i}=C_{\alpha}\kappa^{k}$ and $l^{k}_{\beta,i}=C_{\beta}\kappa^{k}$, with $(1-\gamma)< \kappa <1$ and two constants $C_{\alpha}$ and $C_{\beta}$ as the initial intervals. We know that if for every $k$, the values $x^{k}_{i}$ and $\nabla f_i$ fall inside the quantization intervals, the quantization errors converge at the same linear rate with the constant $\kappa$. In Section~\ref{se:Inexact distributed algorithm with quantization refinement}, we will show that by properly choosing the number of bits $n$ and the initial intervals $C_{\alpha}$ and $C_{\beta}$, it can be guaranteed that $x^{k}_{i}$ and $\nabla f_i$ fall inside the quantization intervals at every iteration and the quantization errors decrease linearly. 

 We add an extra re-projection step (green box) into the algorithm, because the quantized value $\hat{x}^{k}_{\mathcal{N}_i}$ can be an infeasible solution with respect to the constraints $\mathbb{C}_{\mathcal{N}_i}$. The re-projection step guarantees that at each iteration the gradient is computed based on a feasible solution. Using the convexity of the constraints, we can show that the error caused by the re-projected point $\tilde{x}^{k}_{\mathcal{N}_i} = \mbox{Proj}_{\mathbb{C}_{\mathcal{N}_i}} (\hat{x}^{k}_{\mathcal{N}_i})$ is upper-bounded by the quantization error. To summarize, all the errors induced by the limited communication in the distributed algorithm are upper bounded by a linearly decreasing function with the constant $\kappa$, which implies that the distributed algorithm with quantization converges to the global optimum and the linear convergence rate is preserved. These results will be shown in detail in Section~\ref{se:Inexact distributed algorithm with quantization refinement}.

\subsection{Distributed algorithm with quantization refinement}\label{se:Inexact distributed algorithm with quantization refinement}

In this section, we propose a distributed algorithm with a progressive quantization design in Algorithm~\ref{al:PGM for distributed optimization with quantization}. For every sub-system $i$, there are two uniform quantizers $Q^{k}_{\alpha,i}$ and $Q^{k}_{\beta,i}$ using the formulation introduced in Section~\ref{se:uniform quantizer} with a fixed number of bits $n$, changing quantization intervals $l^{k}_{\alpha,i}$ and $l^{k}_{\beta,i}$ and changing mid-values $\bar{x}^{k}_{\alpha,i}$ and $\bar{\nabla} f^{k}_{\beta,i}$ for transmitting $x^{k}_i$, and $\nabla f^{k}_i$ at every iteration $k$. At iteration $k$, the quantization intervals are set to be $l^{k}_{\alpha,i}=C_{\alpha}\kappa^{k}$ and $ l^{k}_{\beta,i}=C_{\beta}\kappa^{k}$, and the mid-values are set to be the previous quantized values $\bar{x}^{k}_{\alpha,i} = \hat{x}^{k-1}_{i}$ and $\bar{\nabla} f^{k}_{\beta,i} = \hat{\nabla} f^{k-1}_i$. The two parameters $C_{\alpha}=l^{0}_{\alpha,i}$ and $C_{\beta}=l^{0}_{\beta,i}$ denote the initial quantization intervals.

In this paper, $\hat{\cdot}$ is used to denote a quantized value, e.g. $\hat{x}^{k}_{i} = Q^{k}_{\alpha,i}(x^{k}_{i})$ and $\tilde{\cdot}$ is used to denote a re-projected value, e.g. $\tilde{x}^{k}_{\mathcal{N}_i} = \mbox{Proj}_{\mathbb{C}_{\mathcal{N}_i}} (\hat{x}^{k}_{\mathcal{N}_i})$. The quantization errors are denoted by $\alpha^{k}_{i}=\hat{x}^{k}_{i}-x^{k}_{i}$ and $\beta^{k}_i=\hat{\nabla} f^{k}_i-\nabla f^{k}_i$.
\begin{scriptsize}
\begin{algorithm}
\caption{Distributed algorithm with quantization refinement}
\begin{algorithmic} 
\REQUIRE Initialize $\hat{x}^{-1}_{i} = x^{0}_{i} = 0$, $\hat{\nabla} f^{-1}_i = \nabla f_i(\mbox{Proj}_{\mathbb{C}_{\mathcal{N}_i}} (0))$, $(1-\gamma)< \kappa <1$ and $\tau < \frac{1}{L}$. 
\FOR {$k=0,1,2,\cdots$}
\STATE For sub-system $i$, $i = 1,2,\cdots ,M$ do in parallel:
\STATE 1: Update the parameters of quantizer $Q^{k}_{\alpha,i}$: $l^{k}_{\alpha,i}=C_{\alpha}\kappa^{k}$ and $\bar{x}^{k}_{\alpha,i} = \hat{x}^{k-1}_{i}$
\STATE 2: Quantize the local variable: $\hat{x}^{k}_{i} = Q^{k}_{\alpha,i}(x^{k}_{i}) = x^{k}_{i} + \alpha^{k}_i$
\STATE 3: Send $\hat{x}^{k}_{i}$ to all the neighbours of sub-system $i$
\STATE 4: Compute the projection of $\hat{x}^{k}_{\mathcal{N}_i}$: $\tilde{x}^{k}_{\mathcal{N}_i} = \mbox{Proj}_{\mathbb{C}_{\mathcal{N}_i}} (\hat{x}^{k}_{\mathcal{N}_i})$
\STATE 5: Compute $\nabla f^{k}_i=\nabla f_i(\tilde{x}^{k}_{\mathcal{N}_i})$
\STATE 6: Update the parameters of quantizer $Q^{k}_{\beta,i}$: $ l^{k}_{\beta,i}=C_{\beta}\kappa^{k}$ and $\bar{\nabla} f^{k}_{\beta,i} = \hat{\nabla} f^{k-1}_i$
\STATE 7: Quantize the gradient: $\hat{\nabla} f^{k}_i = Q^{k}_{\beta,i}(\nabla f^{k}_i) = \nabla f^{k}_i + \beta^{k}_i$
\STATE 8: Send $\hat{\nabla} f^{k}_i$ to all the neighbours of sub-system $i$
\STATE 9: Update the local variable: $ x^{k+1}_{i} =  \mbox{Proj}_{\mathbb{C}_{i}}(x^{k}_{i}-\tau\sum_{j\in\mathcal{N}_i} F_{ji}\hat{\nabla} f^{k}_{j})$
\ENDFOR
\end{algorithmic}
\label{al:PGM for distributed optimization with quantization}
\end{algorithm}\begin{footnotesize}
\end{footnotesize}
\end{scriptsize}

\begin{remark}
We want to highlight Step~4 in Algorithm~\ref{al:PGM for distributed optimization with quantization}, because it is the key step that allows us to extend the algorithm in \cite{pu_quantization_2015} for solving an unconstrained distributed optimization problem to constrained problems. The re-projection step ensures that the point used to compute the gradient at each iteration is a feasible solution subject to the constraints $\mathbb{C}_{\mathcal{N}_i}$, which is a necessary condition for the convergence of the algorithm.
\end{remark}

In the following, we present four lemmas that link Algorithm~\ref{al:PGM for distributed optimization with quantization} to the inexact PGM and prove that Algorithm~\ref{al:PGM for distributed optimization with quantization} converges linearly to the global optimum despite the quantization errors. Lemma~\ref{le:proj(v) is closer to mu than v, for mu in C} states that due to the fact that the constraints are convex, the error between the re-projected point and the original point $\|\tilde{x}^{k}_{\mathcal{N}_i} - x^{k}_{\mathcal{N}_i}\| \leq \|\hat{x}^{k}_{\mathcal{N}_i} - x^{k}_{\mathcal{N}_i}\|$ is upper-bounded by the quantization error. Lemma~\ref{le:ek in the distributed optimization algorithm} shows that the inexactness resulting from quantization in Algorithm~\ref{al:PGM for distributed optimization with quantization} can be considered as the error in the gradient calculation $\{e^{k}\}$ and the error in the computation of the proximal minimization $\{\epsilon^{k}\}$ in Algorithm~\ref{al:inexact ISTA}. Lemma~\ref{le:linear decreasing rate of ek and xk} states that if at each iteration the values $x^{k}_i$ and $\nabla f^{k}_i$ fall inside the quantization intervals, then the errors caused by quantization decrease linearly and the algorithm converges to the global optimum at the same rate. Lemma~\ref{le: conditions on n and quantization intervals} gives conditions on the number of bits and the initial quantization intervals, which guarantee that $x^{k}_i$ and $\nabla f^{k}_i$ fall inside the quantization intervals for each iteration. Once we prove the three lemmas, we are ready to present the main result in Theorem~\ref{th: convergence and complexity bound for the distributed optimization algorithm with quantization}. 

\begin{lemma}\label{le:proj(v) is closer to mu than v, for mu in C}
Let $\mathbb{C}$ be a convex subset of $\mathbb{R}^{n_v}$ and $\mu\in\mathbb{C}$. For any point $v\in \mathbb{R}^{n_v}$, the following holds:
\begin{equation}\label{eq:proj(v) is closer to mu than v, for mu in C}
\|\mu-\mbox{Proj}_{\mathbb{C}}(v)\|\leq \|\mu-v\|\enspace .
\end{equation}
\end{lemma}
\begin{IEEEproof}
Since $\mu\in\mathbb{C}$, we have $\mbox{Proj}_{\mathbb{C}}(\mu) = \mu$. Lemma~\ref{le:proj(v) is closer to mu than v, for mu in C} follows directly from Proposition~2.2.1 in \cite{Bertsekas_Convex_2003}.
\end{IEEEproof}
\begin{lemma}\label{le:ek in the distributed optimization algorithm}
Algorithm~\ref{al:PGM for distributed optimization with quantization} is equivalent to applying the inexact proximal-gradient method in Algorithm~\ref{al:inexact ISTA} to Problem~\ref{pr:distributed optimization} with $\phi = \sum^{M}_{i=1} f_{i}(x_{\mathcal{N}_i})$, $\psi = \sum^{M}_{i=1} I_{\mathbb{C}_i}(x_i)$, 
\begin{equation*}\label{eq:ek in the distributed optimization algorithm}
e^{k} = \sum^{M}_{i = 1} E^{T}_i\nabla f_i(\tilde{x}^{k}_{\mathcal{N}_{i}}) + \sum^{M}_{i = 1} E^{T}_i\beta^{k}_{i} - \sum^{M}_{i = 1} E^{T}_i\nabla f_i(x^{k}_{\mathcal{N}_{i}})\enspace ,
\end{equation*}
and $\epsilon^{k} = \frac{1}{2}\|x^{k}-\tilde{x}^{k}\|^{2}$. Furthermore, $\|e^{k}\|$ and  $\sqrt{\epsilon^{k}}$ are upper-bounded by
\begin{equation}
\|e^{k}\|\leq \sum^{M}_{i=1} L_i\cdot\sum_{j\in\mathcal{N}_{i}}\|\alpha^{k}_j\| + \sum^{M}_{i=1}\|\beta^{k}_i\|\enspace ,
\end{equation}
and
\begin{equation}
\sqrt{\epsilon^{k}}\leq \frac{\sqrt{2}}{2}\sum^{M}_{i=1} \|\alpha^{k}_i\|\enspace .
\end{equation}
\end{lemma}

 The proof of Lemma~\ref{le:ek in the distributed optimization algorithm} will be provided in the appendix in Section~\ref{SEC: appendix proof of Lemma of ek in the distributed optimization algorithm}.

\begin{remark}
Lemma~\ref{le:ek in the distributed optimization algorithm} shows that the errors $\|e^{k}\|$ and $\sqrt{\epsilon^{k}}$ are upper-bounded by functions of the quantization errors $\|\alpha^{k}_i\|$ and $\|\beta^{k}_i\|$. We want to emphasize that the quantization errors $\|\alpha^{k}_i\|$ and $\|\beta^{k}_i\|$ are not necessarily bounded by a linear function with the rate $\kappa$. They are bounded only if the values $x^{k}_{i}$ and $\nabla f_i$ fall inside the quantization intervals that are decreasing at a linear rate. Otherwise, the quantization errors $\|\alpha^{k}_i\|$ and $\|\beta^{k}_i\|$ can be arbitrarily large.
\end{remark}

 From the discussion in Section~\ref{se:inexact PGM}, we know that if $\|e^{k}\|$ and $\sqrt{\epsilon^{k}}$ decrease linearly at a rate larger than $(1-\gamma)$, then $\|x^{k}-x^{\star}\|$ converges linearly at the same rate as $\|e^{k}\|$. Lemma~\ref{le:linear decreasing rate of ek and xk} provides the first step towards achieving this goal. It shows that if the values of $x^{k}_i$ and $\nabla f^{k}_i$ always fall inside the quantization interval, then the computational error of the gradient $\|e^{k}\|$ and the computational error of the proximal operator $\sqrt{\epsilon^{k}}$ as well as $\|x^{k}-x^{\star}\|$ decrease linearly with the constant $\kappa$.
\begin{lemma}\label{le:linear decreasing rate of ek and xk}
For any parameter $\kappa$ satisfying $(1-\gamma)<\kappa < 1$ and a $k\geq 0$, if for all $0\leq p \leq k$ the values of $x^{p}_i$ and $\nabla f^{p}_i$ generated by Algorithm~\ref{al:PGM for distributed optimization with quantization} fall inside of the quantization intervals of $Q^{p}_{\alpha,i}$ and $Q^{p}_{\beta,i}$, i.e. $\|x^{p}_i-\bar{x}^{p}_{\alpha,i}\|_{\infty}\leq \frac{l^{p}_{\alpha,i}}{2}$ and $\|\nabla f^{p}_i-\bar{\nabla} f^{p}_{\beta,i}\|_{\infty}\leq \frac{l^{p}_{\beta,i}}{2}$,  then the error sequences $\|e^{p}\|$ and $\sqrt{\epsilon^{p}}$ satisfy
\begin{equation}\label{eq:bound on ek}
\|e^{p}\| \leq C_1\kappa^{p}\enspace ,\quad \sqrt{\epsilon^{p}} \leq C_2\kappa^{p}\enspace ,
\end{equation}
where $C_1=\frac{M\sqrt{\bar{m}}(L_{max}dC_{\alpha}+\sqrt{d}C_{\beta})}{2^{n+1}}$ and $C_2=\frac{\sqrt{2}}{2}\cdot\frac{M\sqrt{\bar{m}}C_{\alpha}}{2^{n+1}}$, and $\|x^{p+1}-x^{\star}\|$ satisfies
\begin{equation}\label{eq:bound on xk}
\|x^{p+1}-x^{\star}\| \leq  \kappa^{p+1}\left[\|x^{0}-x^{\star}\| + \frac{(C_1+\sqrt{2L}C_2)\kappa}{L(\kappa+\gamma-1)(1-\gamma)}\right]\enspace .
\end{equation}
\end{lemma}
 The proof of Lemma~\ref{le:linear decreasing rate of ek and xk} will be provided in the appendix in Section~\ref{SEC: appendix proof of Lemma of linear decreasing rate of ek and xk}. From Lemma~\ref{le:linear decreasing rate of ek and xk}, we know that the last missing piece is to show that the values $x^{k}_i$ and $\nabla f^{k}_i$ fall inside the quantization interval at every iteration $k$. The following assumption presents conditions on the number of bits $n$ and the initial quantization intervals $C_{\alpha}$ and $C_{\beta}$, which guarantee that for each iteration $x^{k}_i$ and $\nabla f^{k}_i$ in Algorithm~\ref{al:PGM for distributed optimization with quantization} fall inside the changing quantization intervals and the quantization errors decrease linearly with the constant $\kappa$, which further implies that the Algorithm~\ref{al:PGM for distributed optimization with quantization} converges to the global optimum linearly with the same rate $\kappa$. 
\begin{assumption}\label{as: conditions on initial intervals and the number of bits}
Consider the quantizers $Q^{k}_{\alpha,i}$ and $Q^{k}_{\beta,i}$ in Algorithm~\ref{al:PGM for distributed optimization with quantization}. We assume that the parameters of the quantizers, i.e. the number of bits $n$ and the initial quantization intervals $C_{\alpha}$ and $C_{\beta}$ satisfy
\begin{equation}\label{eq:first condition on n and quantization intervals}
a_1 + a_2\frac{C_{\alpha}}{2^{n+1}}+a_3\frac{C_{\beta}}{2^{n+1}} \leq \frac{C_{\alpha}}{2} 
\end{equation}
\begin{equation}\label{eq:second condition on n and quantization intervals}
b_1 + b_2\frac{C_{\alpha}}{2^{n+1}}+b_3\frac{C_{\beta}}{2^{n+1}} \leq \frac{C_{\beta}}{2}\enspace ,
\end{equation}
with
\begin{align*}
& a_1=\frac{(\kappa +1)\|x^{0}-x^{\star}\|}{\kappa}\enspace ,\quad
 a_2= \frac{M\sqrt{\bar{m}}\kappa(\kappa +1)(dL_{\max}+\sqrt{L})+M\sqrt{\bar{m}}L(\kappa+\gamma-1)(1-\gamma)}{L\kappa(\kappa+\gamma-1)(1-\gamma)} \enspace ,\quad
a_3=\frac{M\sqrt{d\bar{m}}(\kappa+1)}{L(\kappa+\gamma-1)(1-\gamma)} \enspace , \\
& b_1= \frac{L_{\max}(\kappa +1)\|x^{0}-x^{\star}\|}{\kappa}\enspace ,\quad
b_2= \frac{L_{\max}M\sqrt{\bar{m}}\kappa(\kappa+1)(dL_{max}+\sqrt{L})+L_{max}d\sqrt{\bar{m}}L(\kappa+1)(\kappa+\gamma-1)(1-\gamma)}{L\kappa(\kappa +\gamma-1)(1-\gamma)}\enspace ,\\
& b_3= \frac{L_{\max}M\sqrt{d\bar{m}}\kappa(\kappa +1)+L\sqrt{d\bar{m}}(\kappa +\gamma -1)(1-\gamma)}{L\kappa(\kappa+\gamma-1)(1-\gamma)}\enspace .
\end{align*}
\end{assumption}
\begin{remark}
The parameters of the quantizers $n$, $C_{\alpha}$ and $C_{\beta}$ are all positive constants. Assumption~\ref{as: conditions on initial intervals and the number of bits} can always be satisfied by increasing $n$, $C_{\alpha}$ and $C_{\beta}$.
\end{remark}
\begin{remark}
For a fixed $n$, inequalities (\ref{eq:first condition on n and quantization intervals}) and (\ref{eq:second condition on n and quantization intervals}) represent two polyhedral constraints on $C_{\alpha}$ and $C_{\beta}$. Therefore, the minimal $C_{\alpha}$ and $C_{\beta}$ can be computed by solving a simple LP problem, i.e. minimizing $C_{\alpha}+C_{\beta}$ subject to $C_{\alpha}\geq 0$, $C_{\beta}\geq 0$, and inequalities (\ref{eq:first condition on n and quantization intervals}) and (\ref{eq:second condition on n and quantization intervals}). Since the minimal $n$ is actually the minimal one guaranteeing that the LP problem has a feasible solution, the minimal $n$ can be found by testing feasibility of the LP problem.
\end{remark}
\begin{lemma}\label{le: conditions on n and quantization intervals}
If Assumption~\ref{as: conditions on initial intervals and the number of bits} is satisfied and $(1-\gamma)<\kappa < 1$, then for any $k\geq 0$ the values of $x^{k}_i$ and $\nabla f^{k}_i$ in Algorithm~\ref{al:PGM for distributed optimization with quantization} fall inside of the quantization intervals of $Q^{k}_{\alpha,i}$ and $Q^{k}_{\beta,i}$, i.e. $\|x^{k}_i-\bar{x}^{k}_{\alpha,i}\|_{\infty}\leq \frac{l^{k}_{\alpha,i}}{2}$ and $\|\nabla f^{k}_i-\bar{\nabla} f^{k}_{\beta,i}\|_{\infty}\leq \frac{l^{k}_{\beta,i}}{2}$.
\end{lemma}

 The proof of Lemma~\ref{le: conditions on n and quantization intervals} will be provided in the appendix in Section~\ref{SEC: appendix proof of Lemma of conditions on n and quantization intervals}. After showing Lemma~\ref{le:ek in the distributed optimization algorithm}, Lemma~\ref{le:linear decreasing rate of ek and xk} and Lemma~\ref{le: conditions on n and quantization intervals}, we are ready to present the main theorem.
\begin{theorem}\label{th: convergence and complexity bound for the distributed optimization algorithm with quantization}
If Assumptions~\ref{as: assumption on f}, \ref{as: assumption on fi} and \ref{as: conditions on initial intervals and the number of bits} hold and $(1-\gamma)<\kappa < 1$, then for $k\geq 0$ the sequence $\{x^{k}\}$ generated by Algorithm~\ref{al:PGM for distributed optimization with quantization} converges to the optimum linearly with the constant $\kappa$ and satisfies
\begin{equation}\label{eq:bound on xk in theorem}
\|x^{k+1}-x^{\star}\| \leq  \kappa^{k+1}\left[\|x^{0}-x^{\star}\| + \frac{(C_1+\sqrt{2L}C_2)\kappa}{L(\kappa+\gamma-1)(1-\gamma)}\right]\enspace .
\end{equation}
with $C_1=\frac{M\sqrt{\bar{m}}(L_{max}dC_{\alpha}+\sqrt{d}C_{\beta})}{2^{n+1}}$ and $C_2=\frac{\sqrt{2}}{2}\cdot\frac{M\sqrt{\bar{m}}C_{\alpha}}{2^{n+1}}$.
\end{theorem}
\begin{IEEEproof}
Since Assumption~\ref{as: assumption on f}, \ref{as: assumption on fi} and \ref{as: conditions on initial intervals and the number of bits} hold, Lemma~\ref{le: conditions on n and quantization intervals} states that for each iteration the values $x^{k}_i$ and $\nabla f^{k}_i$ in Algorithm~\ref{al:PGM for distributed optimization with quantization} fall inside of the quantization intervals of $Q^{k}_{\alpha,i}$ and $Q^{k}_{\beta,i}$. Then from Lemma~\ref{le:linear decreasing rate of ek and xk}, we know that the error sequences $\|e^{k}\|$ and $\sqrt{\epsilon^{k}}$ satisfy $\|e^{k}\| \leq C_1\kappa^{k}$ and $\sqrt{\epsilon^{k}} \leq C_2\kappa^{k}$, and by Lemma~\ref{le:ek in the distributed optimization algorithm} the sequence $x^{k}$ generated by Algorithm~\ref{al:PGM for distributed optimization with quantization} satisfies inequality~(\ref{eq:bound on xk in theorem}).
\end{IEEEproof}

 Recalling the complexity bound in Proposition~\ref{pr:convergence rate of inexact ISTA}, we know that for the case without errors the algorithm converges linearly with the constant $1-\gamma$. After imposing quantization on the algorithm, it still converges to the global optimum linearly but with a larger constant $\kappa >1-\gamma$. We conclude that with the proposed quantization design, the linear convergence of the algorithm is preserved, but the constant of the convergence rate has to be enlarged in order to compensate for the deficiencies from limited communication.

\subsection{Accelerated distributed algorithm with quantization refinement}

In this section, we propose an accelerated variant of the distributed algorithm with quantization refinement in Algorithm~\ref{al:APGM for distributed optimization with quantization} based on the inexact accelerated proximal gradient method in Algorithm~\ref{al:inexact APGM}. Compared to Algorithm~\ref{al:PGM for distributed optimization with quantization}, Algorithm~\ref{al:APGM for distributed optimization with quantization} has an extra accelerating Step~5 $\tilde{y}^{k}_{\mathcal{N}_i} = \tilde{x}^{k}_{\mathcal{N}_i} + \frac{1-\sqrt{\gamma}}{1+\sqrt{\gamma}}(\tilde{x}^{k}_{\mathcal{N}_i}-\tilde{x}^{k-1}_{\mathcal{N}_i})$, and at each iteration the gradient $\nabla f^{k}_i$ is computed based on $\tilde{y}^{k}_{\mathcal{N}_{i}}$. The accelerating step improves the constant of the linear convergence rate of the algorithms from $1-\gamma$ to $\sqrt{1-\sqrt{\gamma}}$, and changes the condition on the quantization parameter $\kappa$ to $\sqrt{1-\sqrt{\gamma}}< \kappa <1$.

\begin{scriptsize}
\begin{algorithm}
\caption{Accelerated distributed algorithm with quantization refinement}
\begin{algorithmic} 
\REQUIRE Initialize $\hat{x}^{-1}_{i} =x^{-1}_{i} =x^{0}_{i} = 0$, $\tilde{x}^{-1}_{\mathcal{N}_i}=0$, $\hat{\nabla} f^{-1}_i = \nabla f_i(\mbox{Proj}_{\mathbb{C}_{\mathcal{N}_i}} (0))$, $\sqrt{1-\sqrt{\gamma}}< \kappa <1$ and $\tau < \frac{1}{L}$. 
\FOR {$k=0,1,2,\cdots$}
\STATE For sub-system $i$, $i = 1,2,\cdots ,M$ do in parallel:
\STATE 1: Update the parameters of quantizer $Q^{k}_{\alpha,i}$: $l^{k}_{\alpha,i}=C_{\alpha}\kappa^{k}$ and $\bar{x}^{k}_{\alpha,i} = \hat{x}^{k-1}_{i}$
\STATE 2: Quantize the local variable: $\hat{x}^{k}_{i} = Q^{k}_{\alpha,i}(x^{k}_{i}) = x^{k}_{i} + \alpha^{k}_i$
\STATE 3: Send $\hat{x}^{k}_{i}$ to all the neighbours of sub-system $i$
\STATE 4: Compute the projection of $\hat{x}^{k}_{\mathcal{N}_i}$: $\tilde{x}^{k}_{\mathcal{N}_i} = \mbox{Proj}_{\mathbb{C}_{\mathcal{N}_i}} (\hat{x}^{k}_{\mathcal{N}_i})$
\STATE 5: Accelerating update: $\tilde{y}^{k}_{\mathcal{N}_i} = \tilde{x}^{k}_{\mathcal{N}_i} + \frac{1-\sqrt{\gamma}}{1+\sqrt{\gamma}}(\tilde{x}^{k}_{\mathcal{N}_i}-\tilde{x}^{k-1}_{\mathcal{N}_i})$ and $y^{k}_{i} = x^{k}_{i} + \frac{1-\sqrt{\gamma}}{1+\sqrt{\gamma}}(x^{k}_{i}-x^{k-1}_{i})$ 
\STATE 6: Compute $\nabla f^{k}_i=\nabla f_i(\tilde{y}^{k}_{\mathcal{N}_i})$
\STATE 7: Update the parameters of quantizer $Q^{k}_{\beta,i}$: $ l^{k}_{\beta,i}=C_{\beta}\kappa^{k}$ and $\bar{\nabla} f^{k}_{\beta,i} = \hat{\nabla} f^{k-1}_i$
\STATE 8: Quantize the gradient: $\hat{\nabla} f^{k}_i = Q^{k}_{\beta,i}(\nabla f^{k}_i) = \nabla f^{k}_i + \beta^{k}_i$
\STATE 9: Send $\hat{\nabla} f^{k}_i$ to all the neighbours of sub-system $i$
\STATE 10: Update the local variable: $ x^{k+1}_{i} =  \mbox{Proj}_{\mathbb{C}_{i}}(y^{k}_{i}-\tau\sum_{j\in\mathcal{N}_i} F_{ji}\hat{\nabla} f^{k}_{j})$
\ENDFOR
\end{algorithmic}
\label{al:APGM for distributed optimization with quantization}
\end{algorithm}\begin{footnotesize}
\end{footnotesize}
\end{scriptsize}

\begin{lemma}\label{le:ek in the accelerated distributed optimization algorithm}
Algorithm~\ref{al:APGM for distributed optimization with quantization} is equivalent to applying the inexact accelerated proximal-gradient method in Algorithm~\ref{al:inexact APGM} to Problem~\ref{pr:distributed optimization} with $\phi = \sum^{M}_{i=1} f_{i}(x_{\mathcal{N}_i})$, $\psi = \sum^{M}_{i=1} I_{\mathbb{C}_i}(x_i)$, 
\begin{equation*}\label{eq:ek in the distributed optimization algorithm}
e^{k} = \sum^{M}_{i = 1} E^{T}_i\nabla f_i(\tilde{y}^{k}_{\mathcal{N}_{i}}) + \sum^{M}_{i = 1} E^{T}_i\beta^{k}_{i} - \sum^{M}_{i = 1} E^{T}_i\nabla f_i(y^{k}_{\mathcal{N}_{i}})\enspace ,
\end{equation*}
and $\epsilon^{k} = \frac{1}{2}\|x^{k}-\tilde{x}^{k}\|^{2}$. Furthermore, $\|e^{k}\|$ and  $\sqrt{\epsilon^{k}}$ are upper-bounded by
\begin{equation}
\|e^{k}\|\leq \sum^{M}_{i=1} L_i\cdot\sum_{j\in\mathcal{N}_{i}}(\frac{2}{1+\sqrt{\gamma}}\|\alpha^{k}_j\|+ \frac{1-\sqrt{\gamma}}{1+\sqrt{\gamma}}\|\alpha^{k-1}_j\|) + \sum^{M}_{i=1}\|\beta^{k}_i\|\enspace .
\end{equation}
and
\begin{equation}
\sqrt{\epsilon^{k}}\leq \frac{\sqrt{2}}{2}\sum^{M}_{i=1} \|\alpha^{k}_i\|\enspace .
\end{equation}
\end{lemma}
\begin{IEEEproof}
The proof follows the same flow of the proof of Lemma~\ref{le:ek in the distributed optimization algorithm}. The only difference is that at each iteration the gradient $\nabla f^{k}_i$ is computed based on $\tilde{y}^{k}_{\mathcal{N}_{i}}$, which is a linear combination of $\tilde{x}^{k}_{\mathcal{N}_{i}}$ and $\tilde{x}^{k-1}_{\mathcal{N}_{i}}$. Hence, the upper-bound on the computational error of the gradient $\|e^{k}\|$ is a function of the linear combination of $\|\alpha^{k-1}_i\|$, $\|\alpha^{k}_i\|$ and $\|\beta^{k}_i\|$.
\end{IEEEproof}

\begin{lemma}\label{le:linear decreasing rate of ek and xk for APGM}
For any parameter $\kappa$ satisfying $\sqrt{1-\sqrt{\gamma}}<\kappa < 1$ and a $k \geq 0$, if for all $0\leq p \leq k$ the values of $x^{p}_i$ and $\nabla f^{p}_i$ generated by Algorithm~\ref{al:APGM for distributed optimization with quantization} fall inside of the quantization intervals of $Q^{p}_{\alpha,i}$ and $Q^{p}_{\beta,i}$, i.e. $\|x^{k}_i-\bar{x}^{k}_{\alpha,i}\|_{\infty}\leq \frac{l^{k}_{\alpha,i}}{2}$ and $\|\nabla f^{k}_i-\bar{\nabla} f^{k}_{\beta,i}\|_{\infty}\leq \frac{l^{k}_{\beta,i}}{2}$,  then the sequences $\|e^{p}\|$ and $\sqrt{\epsilon^{p}}$ satisfy
\begin{equation}\label{eq:bound on ek}
\|e^{p}\| \leq C_3\kappa^{p}\enspace ,\quad \sqrt{\epsilon^{p}} \leq C_4\kappa^{p}\enspace.
\end{equation}
where $C_3=\frac{M\sqrt{\bar{m}}(3L_{max}dC_{\alpha}+\kappa \sqrt{d}C_{\beta})}{\kappa\cdot 2^{n+1}}$ and $C_4=\frac{\sqrt{2}}{2}\cdot\frac{M\sqrt{\bar{m}}C_{\alpha}}{2^{n+1}}$, and $\|x^{p+1}-x^{\star}\|$ satisfies
\begin{equation}
\|x^{p+1}-x^{\star}\| \leq  \kappa^{p+1}\left[\frac{2\sqrt{\Phi(x^{0})-\Phi(x^{\star})}}{\sqrt{\sigma_{\phi}}} + \frac{(2C_3+2\sqrt{2L}C_4+\sqrt{2\sigma_{\phi}}C_4)\kappa}{\sigma_{\phi}(\kappa-\sqrt{1-\sqrt{\gamma}})\cdot \sqrt{1-\sqrt{\gamma}}}\right]\enspace .
\end{equation}
\end{lemma}

\begin{IEEEproof}
The proof follows the same flow of the proof of Lemma~\ref{le:linear decreasing rate of ek and xk} by replacing the upper-bounds on $\|e^{k}\|$ and $\sqrt{\epsilon^{k}}$ in Lemma~\ref{le:ek in the distributed optimization algorithm} and the upper-bound on $\|x^{p+1}-x^{\star}\|$ in Proposition~\ref{pr:convergence rate of inexact ISTA} by the ones in Lemma~\ref{le:ek in the accelerated distributed optimization algorithm} and Proposition~\ref{pr:convergence rate of inexact APGM}. In addition, the proof requires the fact that $\sqrt{1-\sqrt{\gamma}}< \kappa <1$ and $1< 1+\sqrt{\gamma}< 2$.
\end{IEEEproof}

\begin{assumption}\label{as: conditions on initial intervals and the number of bits for the accelerated algorithm}
We assume that the number of bits $n$ and the initial quantization intervals $C_{\alpha}$ and $C_{\beta}$ satisfy
\begin{equation}\label{eq:first condition on n and quantization intervals for accelerated case}
a_4 + a_5\frac{C_{\alpha}}{2^{n+1}}+a_6\frac{C_{\beta}}{2^{n+1}} \leq \frac{C_{\alpha}}{2} 
\end{equation}
\begin{equation}\label{eq:second condition on n and quantization intervals for accelerated case}
b_4 + b_5\frac{C_{\alpha}}{2^{n+1}}+b_6\frac{C_{\beta}}{2^{n+1}} \leq \frac{C_{\beta}}{2}\enspace ,
\end{equation}
with
\begin{align*}
& a_4=\frac{2(\kappa +1)\sqrt{\Phi(x^{0})-\Phi(x^{\star})}}{\kappa\sqrt{\sigma_{\phi}}}\enspace ,\\
& a_5= \frac{6M\sqrt{\bar{m}}(\kappa +1)dL_{\max} + M\sqrt{\bar{m}}\kappa(\kappa +1)(2\sqrt{L}+\sqrt{\sigma_{\phi}}) + \sigma_{\phi}M\sqrt{\bar{m}}(\kappa-\sqrt{1-\sqrt{\gamma}})\cdot \sqrt{1-\sqrt{\gamma}}}{\sigma_{\phi}\kappa(\kappa-\sqrt{1-\sqrt{\gamma}})\cdot \sqrt{1-\sqrt{\gamma}}} \enspace ,\\
 &a_6=\frac{2M\sqrt{d\bar{m}}(\kappa+1)}{\sigma_{\phi}(\kappa-\sqrt{1-\sqrt{\gamma}})\cdot \sqrt{1-\sqrt{\gamma}}} \enspace , \\
& b_4= \frac{2L_{\max}(2\kappa^{2} + 3\kappa +1)\sqrt{\Phi(x^{0})-\Phi(x^{\star})}}{\kappa^{2}\sqrt{\sigma_{\phi}}}\enspace ,\\
 & b_5= \frac{L_{max}\sqrt{\bar{m}}(2\kappa^{2} + 3\kappa +1)}{\kappa^{2}}\left[d + \frac{6MdL_{max}+M\kappa(2\sqrt{L} + \sqrt{\sigma_{\phi}})}{\sigma_{\phi}(\kappa-\sqrt{1-\sqrt{\gamma}})\cdot \sqrt{1-\sqrt{\gamma}}}\right]\enspace ,\\
& b_6= \frac{2L_{\max}M\sqrt{d\bar{m}}(2\kappa^{2} + 3\kappa +1)+\sigma_{\phi}\sqrt{d\bar{m}}(\kappa-\sqrt{1-\sqrt{\gamma}})\cdot \sqrt{1-\sqrt{\gamma}}}{\sigma_{\phi}\kappa(\kappa-\sqrt{1-\sqrt{\gamma}})\cdot \sqrt{1-\sqrt{\gamma}}}\enspace .
\end{align*}
\end{assumption}

\begin{lemma}\label{le: conditions on n and quantization intervals for APGM}
If Assumption~\ref{as: conditions on initial intervals and the number of bits for the accelerated algorithm} is satisfied and $\sqrt{1-\sqrt{\gamma}}<\kappa < 1$, then for any $k\geq 0$ the values of $x^{k}_i$ and $\nabla f^{k}_i$ in Algorithm~\ref{al:APGM for distributed optimization with quantization} fall inside of the quantization intervals of $Q^{k}_{\alpha,i}$ and $Q^{k}_{\beta,i}$, i.e. $\|x^{k}_i-\bar{x}^{k}_{\alpha,i}\|_{\infty}\leq \frac{l^{k}_{\alpha,i}}{2}$ and $\|\nabla f^{k}_i-\bar{\nabla} f^{k}_{\beta,i}\|_{\infty}\leq \frac{l^{k}_{\beta,i}}{2}$.
\end{lemma}

 The proof of Lemma~\ref{le: conditions on n and quantization intervals for APGM} will be provided in the appendix in Section~\ref{SEC: appendix proof of Lemma of conditions on n and quantization intervals for APGM}. 

\begin{theorem}\label{th: convergence and complexity bound for the accelerated distributed optimization algorithm with quantization}
If Assumptions~\ref{as: assumption on f}, \ref{as: assumption on fi} and \ref{as: conditions on initial intervals and the number of bits for the accelerated algorithm} hold and $\sqrt{1-\sqrt{\gamma}}<\kappa < 1$, then for $k\geq 0$ the sequence $\{x^{k}\}$ generated by Algorithm~\ref{al:APGM for distributed optimization with quantization} converges to the optimum linearly with the constant $\kappa$ and satisfies
\begin{equation}\label{eq:bound on xk in theorem for APGM}
\|x^{k+1}-x^{\star}\| \leq  \kappa^{k+1}\left[\frac{2\sqrt{\Phi(x^{0})-\Phi(x^{\star})}}{\sqrt{\sigma_{\phi}}} + \frac{(2C_3+2\sqrt{2L}C_4+\sqrt{2\sigma_{\phi}}C_4)\kappa}{\sigma_{\phi}(\kappa-\sqrt{1-\sqrt{\gamma}})\cdot \sqrt{1-\sqrt{\gamma}}}\right]\enspace ,
\end{equation}
with $C_3=\frac{M\sqrt{\bar{m}}(3L_{max}dC_{\alpha}+\kappa \sqrt{d}C_{\beta})}{\kappa\cdot 2^{n+1}}$ and $C_4=\frac{\sqrt{2}}{2}\cdot\frac{M\sqrt{\bar{m}}C_{\alpha}}{2^{n+1}}$.
\end{theorem}
\begin{IEEEproof}
The proof follows directly from the proof of Theorem~\ref{th: convergence and complexity bound for the distributed optimization algorithm with quantization} by replacing Lemma~\ref{le:ek in the distributed optimization algorithm}, Lemma~\ref{le:linear decreasing rate of ek and xk} and Lemma~\ref{le: conditions on n and quantization intervals} by Lemma~\ref{le:ek in the accelerated distributed optimization algorithm}, Lemma~\ref{le:linear decreasing rate of ek and xk for APGM} and Lemma~\ref{le: conditions on n and quantization intervals for APGM}.
\end{IEEEproof}

\section{Numerical Example}\label{SEC:example}
This section illustrates the theoretical findings of the paper and demonstrates the performance of Algorithm~\ref{al:PGM for distributed optimization with quantization} and Algorithm~\ref{al:APGM for distributed optimization with quantization} for solving a distributed quadratic programming (QP) problem originating from the problem of regulating constrained distributed linear systems by model predictive control (MPC) in the form of Problem~\ref{pr:distributed MPC}. For more information about distributed MPC, see e.g. \cite{conte_distributed_2012}, \cite{conte_computational_2012} and \cite{pu_inexact_2014}. 

\begin{problem}\label{pr:distributed MPC}
\begin{align*}
\min_{z,u}  & \quad \sum^{M}_{i=1} \sum^{N-1}_{t=0} l_i(z_{i}(t),u_{i}(t)) + \sum^{M}_{i=1} l^{f}_i(z_{i}(N)) \\
s.t. \quad &z_{i}(t+1) = A_{ii}z_{j}(t) + \sum_{j\in \mathcal{N}_{i}} B_{ij}u_{j}(t)\\
& u_i(t) \in \mathbb{U}_i\enspace ,\; z_i(0) = \bar{z}_{i}\enspace , \; i=1,2,\cdots ,M\enspace ,
\end{align*}
\end{problem}

$M$ and $N$ denote the number of subsystems and the horizon of the MPC problem, respectively. The state and input sequences along the horizon of subsystem $i$ are denoted by $z_{i}=[z^{T}_{i}(0), z^{T}_{i}(1),\cdots, z^{T}_{i}(N)]^{T}$ and $u_{i}=[u^{T}_{i}(0),u^{T}_{i}(1),\cdots,u^{T}_{i}(N-1)]^{T}$. The discrete-time linear dynamics of subsystem $i$ are given by $z_{i}(t+1) = A_{ii}z_{j}(t) + \sum_{j\in \mathcal{N}_{i}} B_{ij}u_{j}(t)$, where $A_{ii}$ and $B_{ij}$ are the dynamic matrices. The initial state is denoted by $\bar{z}_{i}$. The control inputs of subsystem $i$ are subject to local convex constraints $u_i(t) \in \mathbb{U}_i$. $l_i(\cdot,\cdot)$ and $l^{f}_i(\cdot)$ are strictly convex cost functions. From Problem~\ref{pr:distributed MPC}, we can see that subsystem $i$ is coupled with its neighbours in the linear dynamics. 
 
 We randomly generate a distributed MPC problem in the form of Problem~\ref{pr:distributed MPC}. We first randomly generate a connected network with $M=40$ sub-systems. Each sub-system has $3$ states and $2$ inputs. The dynamical matrices $A_{ii}$ and $B_{ij}$ are randomly generated, i.e. generally dense, and the local systems are controllable and unstable. The input constraint $\mathbb{U}_i$ for sub-system $i$ is set to $\mathbb{U}_i = \{u_i| -0.4\cdot\mathbf{1} \leq u_i(t)\leq 0.3\cdot\mathbf{1}\}$, where $\mathbf{1}$ denotes the all-ones vector with the same dimension as $u_i$. The horizon of the MPC problem is set to $N=11$. The local cost functions are chosen as quadratic functions $l_i(z_i(t),u_i(t)) = z^{T}_{i}(t)Q_iz_i(t)+u^{T}_{i}(t)R_iu_i(t)$ and $l^{f}_i(z_i(N)) = z^{T}_{i}(N)P_iz_i(N)$, where $Q_i$, $R_i$ and $P_i$ are identity matrices. The initial states $\bar{z}_{i}$ are chosen, such that more than $50\%$ of the optimization variables are at the constraints at optimality. 
 
\begin{problem}\label{pr:distributed QP}
\begin{align*}
\min_{x\in \mathbb R^{n_x}} \; f(x) =&\sum^{M}_{i=1} f_{i}(x_{\mathcal{N}_{i}}) = \sum^{M}_{i=1} x^{T}_{\mathcal{N}_{i}}H_{i}x_{\mathcal{N}_{i}}+h_{i}x_{\mathcal{N}_{i}} \\
s.t. \quad & x_i \in \mathbb{C}_i\enspace \enspace .
\end{align*}
\end{problem}

 By eliminating all state variables distributed MPC problems of this class can be reformulated as a distributed QP of the form in Problem~\ref{pr:distributed QP} with the local variables $x_i = u_i$ and the concatenations of the variables of subsystem $i$ and its neighbours $x_{\mathcal{N}_{i}}$. Matrix $H_i$ is dense and positive definite, and vector $h_i$ is dense. The constraint $\mathbb{C}_i = \mathbb{U}^{N}_i$ is a polytopic set. 

  Table~\ref{ta:the parameters of the two examples} shows the parameters chosen in Algorithm~\ref{al:PGM for distributed optimization with quantization} and Algorithm~\ref{al:APGM for distributed optimization with quantization}, including the constants of the convergence rate of the algorithms, i.e. $\gamma = \frac{\sigma_f}{L}$ and $\sqrt{1-\sqrt{\gamma}}$, the decrease rates of the quantization intervals $\kappa$ satisfying $1-\gamma\leq \kappa \leq 1$ for Algorithm~\ref{al:PGM for distributed optimization with quantization} and $\sqrt{1-\sqrt{\gamma}}\leq \kappa \leq 1$ for Algorithm~\ref{al:APGM for distributed optimization with quantization} and the minimum number of bits required for convergence $n_{min}$.
 
\begin{figure}[htbp]
   \centering
      \includegraphics[width=0.5\linewidth]{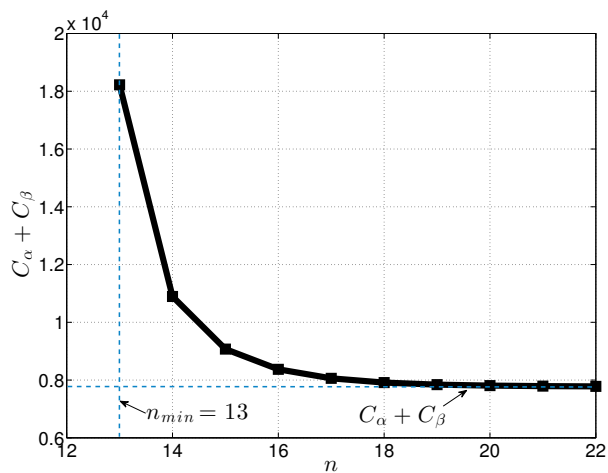}
   \caption{Relationship between the number of bits $n$ and the minimum initial quantization intervals $C_{\alpha}$ and $C_{\beta}$ satisfying Assumption~\ref{as: conditions on initial intervals and the number of bits} for Problem~\ref{pr:distributed QP} originating from the distributed MPC problem.}
   \label{fig:relationship between n and the initial intervals.}
\end{figure}

 Fig.~\ref{fig:relationship between n and the initial intervals.} shows the relationship between the number of bits $n$ and the minimum initial quantization intervals $C_{\alpha}$ and $C_{\beta}$, which satisfy Assumption~\ref{as: conditions on initial intervals and the number of bits} for Problem~\ref{pr:distributed QP}. We see that the minimum number of bits required for convergence is equal to $n_{min}=13$, and as the number of bits $n$ increases, the required minimum $C_{\alpha}$ and $C_{\beta}$ decrease.

\begin{figure}[htbp]
   \centering
      \includegraphics[width=0.5\linewidth]{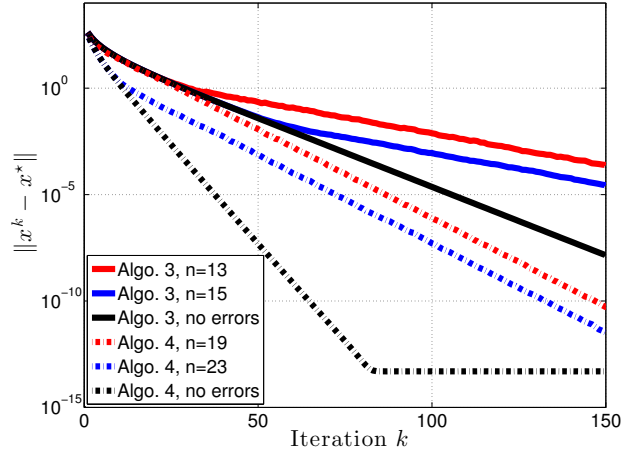}
   \caption{Comparison of the performance of Algorithm~\ref{al:PGM for distributed optimization with quantization} and Algorithm~\ref{al:APGM for distributed optimization with quantization} with different $n$ and corresponding minimum $C_{\alpha}$ and $C_{\beta}$ with the exact algorithms (no quantization errors) for Problem~\ref{pr:distributed QP}.}
   \label{fig: the comparison the algorithm with different quantization parameters for example 1.}
\end{figure}

 Fig.~\ref{fig: the comparison the algorithm with different quantization parameters for example 1.} shows the performance of Algorithm~\ref{al:PGM for distributed optimization with quantization} and Algorithm~\ref{al:APGM for distributed optimization with quantization} for solving the distributed QP problem in Problem~\ref{pr:distributed QP} originating from the distributed MPC problem. For Algorithm~\ref{al:PGM for distributed optimization with quantization}, $n$ is set to $13$ and $15$, respectively, and the initial quantization intervals $C_{\alpha}$ and $C_{\beta}$ are set to corresponding minimum values satisfying Assumption~\ref{as: conditions on initial intervals and the number of bits}. For Algorithm~\ref{al:APGM for distributed optimization with quantization}, $n$ is set to $19$ and $23$, and $C_{\alpha}$ and $C_{\beta}$ to corresponding minimum values satisfying Assumption~\ref{as: conditions on initial intervals and the number of bits for the accelerated algorithm}. In Fig.~\ref{fig: the comparison the algorithm with different quantization parameters for example 1.} we can observe that the proposed distributed algorithms with quantization converges to the global optimum linearly and the performance is improved when the number of bits $n$ is increased. Due to the acceleration step, Algorithm~\ref{al:APGM for distributed optimization with quantization} converges faster than Algorithm~\ref{al:PGM for distributed optimization with quantization}. However, Algorithm~\ref{al:APGM for distributed optimization with quantization} requires a larger number of bits $n$ to guarantee the convergence.
 
\begin{table*}
\centering
\begin{tabular}{|c|c|c|c|c|} \hline
Parameters &  Algorithm~\ref{al:PGM for distributed optimization with quantization}   &  Algorithm~\ref{al:APGM for distributed optimization with quantization}                                 \\ \hline
Constant of rate            &         $1 - \gamma=0.8093$  & $\sqrt{1 - \sqrt{\gamma}}=0.7505$                \\ \hline
$\kappa$   &   $0.9333$                                 & $0.7991$                                                              \\ \hline
$n_{min}$  &    $13$                                        & $19$                                                                     \\ \hline
\end{tabular}
\vspace{0.5em}
\caption{The parameters in Algorithm~\ref{al:PGM for distributed optimization with quantization} and Algorithm~\ref{al:APGM for distributed optimization with quantization} for solving Problem~\ref{pr:distributed QP}.}
\label{ta:the parameters of the two examples}
\end{table*}

\section{Appendix}\label{SEC: appendix}

\subsection{Proof of Proposition~\ref{pr:convergence rate of inexact APGM}}\label{SEC: appendix proof of Proposition of convergence rate of inexact APGM}

\begin{IEEEproof}
By the strong convexity of the function $\phi$, we know
\begin{equation*}
\frac{\sigma_{\phi}}{2}\|x^{k+1}-x^{\star}\|^{2} \leq \Phi (x^{k+1}) - \Phi (x^{\star})\enspace .
\end{equation*}
From Proposition~4 in \cite{schmidt_convergence_2011}, it follows that
\begin{equation*}
\|x^{k+1}-x^{\star}\|^{2} \leq \frac{2}{\sigma_{\phi}}(1-\sqrt{\gamma})^{k+1}\left(\sqrt{2(\Phi (x^{0}) - \Phi (x^{\star}))}+\sqrt{\frac{2}{\sigma_{\Phi}}}\sum^{k}_{p=0} (\|e^{p}\|+\sqrt{2L(\nabla \phi)\epsilon^{p}})(1-\sqrt{\gamma})^{-\frac{p+1}{2}}+\sqrt{\sum^{k}_{p=0}\epsilon^{p}(1-\sqrt{\gamma})^{-p-1}}\right)^{2}\enspace .
\end{equation*}
By the fact $\sqrt{v+\mu}\leq \sqrt{v} +\sqrt{\mu}$ for any $v,\mu\in \mathbb{R}_{+}$, we simplify the inequality above as
\begin{equation*}
\|x^{k+1}-x^{\star}\|^{2} \leq \frac{2}{\sigma_{\phi}}(1-\sqrt{\gamma})^{k+1}\left(\sqrt{2(\Phi (x^{0}) - \Phi (x^{\star}))}+\sqrt{\frac{2}{\sigma_{\phi}}}\sum^{k}_{p=0} (\|e^{p}\|+(\sqrt{2L(\nabla \phi)}+\sqrt{\frac{\sigma_{\phi}}{2}})\sqrt{\epsilon^{p}})(1-\sqrt{\gamma})^{-\frac{p+1}{2}}\right)^{2}\enspace .
\end{equation*}
Taking the square-root of both sides of the inequality above, we get inequality~(\ref{eq:complexity bound for inexact APGM}).
\end{IEEEproof}


\subsection{Proof of Lemma~\ref{le:ek in the distributed optimization algorithm}}\label{SEC: appendix proof of Lemma of ek in the distributed optimization algorithm}

\begin{IEEEproof}
By the definition, the gradient computation error $e^{k}$ in Algorithm~\ref{al:inexact ISTA} is equal to
\begin{align*}
e^{k} 
&= \hat{\nabla}f(\tilde{x}^{k}) - \nabla f(x^{k}) = \sum^{M}_{i = 1} E^{T}_i\hat{\nabla} f_i(\tilde{x}^{k}_{\mathcal{N}_{i}}) -\sum^{M}_{i = 1} E^{T}_i\nabla f_i(x^{k}_{\mathcal{N}_{i}}) \\
& = \sum^{M}_{i = 1} E^{T}_i\nabla f_i(\tilde{x}^{k}_{\mathcal{N}_{i}}) + \sum^{M}_{i = 1} E^{T}_i\beta^{k}_{i} - \sum^{M}_{i = 1} E^{T}_i\nabla f_i(x^{k}_{\mathcal{N}_{i}}).
\end{align*}
Then,
\begin{align*}
\|e^{k}\|\leq \sum^{M}_{i=1} \|E^{T}_i\|\cdot L_i\cdot \|\tilde{x}^{k}_{\mathcal{N}_i} - x^{k}_{\mathcal{N}_i}\| + \sum^{M}_{i=1} \|E^{T}_i\|\|\beta^{k}_i\|\enspace .
\end{align*}
Note that the matrix $E_i$ is a selection matrix, then $\|E^{T}_i\| = 1$. Since $x^{k}_{\mathcal{N}_i}\in\mathbb{C}_{\mathcal{N}_i}$ and $\tilde{x}^{k}_{\mathcal{N}_i} = \mbox{Proj}_{\mathbb{C}_{\mathcal{N}_i}}(\hat{x}^{k}_{\mathcal{N}_i})$, Lemma~\ref{le:proj(v) is closer to mu than v, for mu in C} implies $\|\tilde{x}^{k}_{\mathcal{N}_i} - x^{k}_{\mathcal{N}_i}\|\leq \|\hat{x}^{k}_{\mathcal{N}_i} - x^{k}_{\mathcal{N}_i}\|$. Hence, we have
\begin{align*}
\|e^{k}\|\leq \sum^{M}_{i=1}  L_i\cdot \|\hat{x}^{k}_{\mathcal{N}_i} - x^{k}_{\mathcal{N}_i}\| + \sum^{M}_{i=1} \|\beta^{k}_i\| \leq \sum^{M}_{i=1} L_i\cdot\sum_{j\in\mathcal{N}_{i}}\|\alpha^{k}_j\| + \sum^{M}_{i=1}\|\beta^{k}_i\|\enspace .
\end{align*}
By definition in (\ref{eq:epsilon error in proximity operator}) and the fact that $x^{k}\in\mathbb{C}$ and $\tilde{x}^{k} = \mbox{Proj}_{\mathbb{C}}(\hat{x}^{k})$, we know $\epsilon^{k} = \frac{1}{2}\|x^{k}-\tilde{x}^{k}\|^{2}$. Lemma~\ref{le:proj(v) is closer to mu than v, for mu in C} again implies $\|x^{k}-\tilde{x}^{k}\|\leq \|x^{k}-\hat{x}^{k}\|$. Hence, we have
\begin{align*}
\sqrt{\epsilon^{k}}=\frac{\sqrt{2}}{2}\|x^{k}-\tilde{x}^{k}\|\leq \frac{\sqrt{2}}{2}\|x^{k}-\hat{x}^{k}\|\leq \frac{\sqrt{2}}{2}\sum^{M}_{i=1} \|\alpha^{k}_i\|\enspace .
\end{align*}
\end{IEEEproof}


\subsection{Proof of Lemma~\ref{le:linear decreasing rate of ek and xk}}\label{SEC: appendix proof of Lemma of linear decreasing rate of ek and xk}

\begin{IEEEproof}
From the property of the uniform quantizer, we know that if $x^{p}_i$ and $\nabla f^{p}_i$ fall inside of the quantization intervals of $Q^{p}_{\alpha,i}$ and $Q^{p}_{\beta,i}$, then the quantization errors $\alpha^{p}_{i}$ and $\beta^{p}_{i}$ are upper-bounded by
\begin{align*}
\|\alpha^{p}_{i}\|\leq \sqrt{m_{i}}\cdot\|\alpha^{p}_{i}\|_{\infty} \leq \sqrt{m_{i}}\cdot \frac{l^{p}_{\alpha,i}}{2^{n+1}}\leq \sqrt{\bar{m}}\cdot \frac{l^{p}_{\alpha,i}}{2^{n+1}}\enspace ,\quad
\|\beta^{p}_{i}\| \leq \sqrt{\sum_{j\in\mathcal{N}_i} m_{i}}\cdot\|\beta^{p}_{i}\|_{\infty} \leq \sqrt{\sum_{j\in\mathcal{N}_i} m_{i}}\cdot \frac{l^{p}_{\beta,i}}{2^{n+1}}\leq \sqrt{d\bar{m}}\cdot \frac{l^{p}_{\beta,i}}{2^{n+1}}\enspace ,
\end{align*}
where $\bar{m} := \max_{1\leq i \leq M} m_i$ and $d$ denotes the degree of the graph of the distributed optimization problem. From Lemma~\ref{le:ek in the distributed optimization algorithm}, we have
\begin{align*}
\|e^{p}\| \leq \sum^{M}_{i=1} L_i\cdot\sum_{j\in\mathcal{N}_{i}} \frac{\sqrt{\bar{m}}\cdot l^{p}_{\alpha,j}}{2^{n+1}} + \sum^{M}_{i=1}\frac{\sqrt{d\bar{m}}\cdot l^{p}_{\beta,i}}{2^{n+1}}\enspace ,
\end{align*}
and
\begin{equation*}
\sqrt{\epsilon^{k}}\leq \frac{\sqrt{2}}{2}\sum^{M}_{i=1} \frac{\sqrt{\bar{m}} l^{p}_{\alpha,i}}{2^{n+1}} \enspace .
\end{equation*}
Since the quantization intervals are set to $l^{p}_{\alpha,i}=C_{\alpha}\kappa^{p}$ and $l^{p}_{\beta,i}=C_{\beta}\kappa^{p}$, it implies that
\begin{align*}
\|e^{p}\| 
&\leq \frac{ML_{max}d\sqrt{\bar{m}}\cdot C_{\alpha}\kappa^{p}}{2^{n+1}} + \frac{M\sqrt{d\bar{m}}\cdot C_{\beta}\kappa^{p}}{2^{n+1}}=C_1\kappa^{p}\enspace ,
\end{align*}
and
\begin{equation*}
\sqrt{\epsilon^{k}}\leq \frac{\sqrt{2}}{2}\cdot\frac{M\sqrt{\bar{m}} C_{\alpha}\kappa^{p}}{2^{n+1}}=C_2\kappa^{p} \enspace ,
\end{equation*}
with $C_1=\frac{M\sqrt{\bar{m}}(L_{max}dC_{\alpha}+\sqrt{d}C_{\beta})}{2^{n+1}}$ and $C_2=\frac{\sqrt{2}}{2}\cdot\frac{M\sqrt{\bar{m}}C_{\alpha}}{2^{n+1}}$, where $L_{max} := \max_{1\leq i \leq M} L_i$. Since $(1-\gamma)< \kappa<1$, Lemma~\ref{le:ek in the distributed optimization algorithm} and Proposition~\ref{pr:convergence rate of inexact ISTA} imply that for $0\leq p\leq k$
\begin{align*}
\|x^{p+1}-x^{\star}\| 
& \leq (1-\gamma)^{p+1}\|x^{0}-x^{\star}\| + \frac{(C_1+\sqrt{2L}C_2)}{L}\sum^{p}_{q=0}\kappa^{q}(1-\gamma)^{p+1-q-1}\\
& \leq \kappa^{p+1}\left[\|x^{0}-x^{\star}\| + \frac{(C_1+\sqrt{2L}C_2)}{L(1-\gamma)}\sum^{p}_{q=0}(\frac{1-\gamma}{\kappa})^{p+1-q}\right] .
\end{align*}
Since $0<(1-\gamma)< \kappa <1$, by using the property of geometric series, we get that the expression above is equal to
\begin{align*}
 =  \kappa^{p+1}\left[\|x^{0}-x^{\star}\| + \frac{(C_1+\sqrt{2L}C_2)}{L(1-\gamma)}\cdot\frac{1-(\frac{1-\gamma}{\kappa})^{p+1}}{1-\frac{1-\gamma}{\kappa}}\right] \leq  \kappa^{p+1}\left[\|x^{0}-x^{\star}\| + \frac{(C_1+\sqrt{2L}C_2)\kappa}{L(\kappa+\gamma-1)(1-\gamma)}\right]\enspace .
\end{align*}
Hence, inequality~(\ref{eq:bound on xk}) is proven.
\end{IEEEproof}


\subsection{Proof of Lemma~\ref{le: conditions on n and quantization intervals}}\label{SEC: appendix proof of Lemma of conditions on n and quantization intervals}
\begin{IEEEproof}
We will prove Lemma~\ref{le: conditions on n and quantization intervals} by induction.
\begin{itemize}
\item Base case: When $k = 0$, since $C_{\alpha}$ and $C_{\beta}$ are positive numbers and $\hat{x}^{-1}_{i}$ and $x^{0}_{i}$ are initialized to zero, it holds that $\|x^{0}_i-\bar{x}^{0}_{\alpha,i}\|_{\infty} = \|x^{0}_i-\hat{x}^{-1}_{i} \|_{\infty} = 0 \leq \frac{l^{0}_{\alpha,i}}{2} = \frac{C_{\alpha}}{2}$ and $\|\nabla f^{0}_i-\bar{\nabla} f^{0}_{\beta,i}\|_{\infty}= \|\nabla f^{0}_i-\hat{\nabla} f^{-1}_i\|_{\infty} = \|\nabla f_i(\tilde{x}^{0}_{\mathcal{N}_i})-\nabla f_i(\mbox{Proj}_{\mathbb{C}_{\mathcal{N}_i}} (0))\| = 0 \leq \frac{l^{0}_{\beta,i}}{2}= \frac{C_{\beta}}{2}$.
\item Induction step: Let $g\geq 0$ be given and suppose that $\|x^{k}_i-\bar{x}^{k}_{\alpha,i}\|_{\infty}\leq \frac{l^{k}_{\alpha,i}}{2}$ and $\|\nabla f^{k}_i-\bar{\nabla} f^{k}_{\beta,i}\|_{\infty}\leq \frac{l^{k}_{\beta,i}}{2}$ for $0\leq k\leq g$. We will prove that
\begin{equation}\label{eq: x falls in the interval at g+1}
\|x^{g+1}_i-\bar{x}^{g+1}_{\alpha,i}\|_{\infty}\leq \frac{l^{g+1}_{\alpha,i}}{2}
\end{equation}
and
\begin{equation}\label{eq: the gradient of f falls in the interval at g+1}
\|\nabla f^{g+1}_i-\bar{\nabla} f^{g+1}_{\beta,i}\|_{\infty}\leq \frac{l^{g+1}_{\beta,i}}{2}
\end{equation}
 for $i=1,\cdots,M$. We first show (\ref{eq: x falls in the interval at g+1}). From Algorithm~\ref{al:PGM for distributed optimization with quantization}, we know
\begin{align*}
\|x^{g+1}_i-\bar{x}^{g+1}_{\alpha,i}\|_{\infty}& =\|x^{g+1}_i-\hat{x}^{g}_{i}\|_{\infty}\\
&\leq\|x^{g+1}-\hat{x}^{g}\|_{\infty} \\
&= \|x^{g+1}-x^{g}-\sum^{M}_{i=1} E^{T}_{i}F^{T}_{ii}\alpha^{g}_{i}\|_{\infty}\\ 
&\leq \|x^{g+1}-x^{g}\|_{\infty}+\|\sum^{M}_{i=1} E^{T}_{i}F^{T}_{ii}\alpha^{g}_{i}\|_{\infty}\\ 
&\leq \|x^{g+1}-x^{\star}\|_{\infty}+\|x^{g}-x^{\star}\|_{\infty}+\|\sum^{M}_{i=1} E^{T}_{i}F^{T}_{ii}\alpha^{g}_{i}\|_{\infty}\enspace .
\end{align*}
Since $E_i$ and $F_{ii}$ are selection matrices, then $\|E_i\|=\|F_{ii}\|=1$. The term above is upper-bounded by
\begin{align*}
\leq \|x^{g+1}-x^{\star}\|_{2}+\|x^{g}-x^{\star}\|_{2}+\sum^{M}_{i=1} \|\alpha^{g}_{i}\|_{2}\enspace .
\end{align*}
By the assumption of the induction, we know $\|x^{k}_i-\bar{x}^{k}_{\alpha,i}\|_{\infty}\leq \frac{l^{k}_{\alpha,i}}{2}$ and $\|\nabla f^{k}_i-\bar{\nabla} f^{k}_{\beta,i}\|_{\infty}\leq \frac{l^{k}_{\beta,i}}{2}$ for $0\leq k\leq g$. Then, using Lemma~\ref{le:linear decreasing rate of ek and xk}, we obtain that the term above is upper-bounded by
\begin{align*}
&\leq \kappa^{g+1}\bigg[\|x^{0}-x^{\star}\|+\frac{(C_1+\sqrt{2L}C_2)\kappa}{L(\kappa+\gamma-1)(1-\gamma)}\bigg] +\kappa^{g}\bigg[\|x^{0}-x^{\star}\|+\frac{(C_1+\sqrt{2L}C_2)\kappa}{L(\kappa+\gamma-1)(1-\gamma)}\bigg]+\frac{M\sqrt{\bar{m}}C_{\alpha}\kappa^{g}}{2^{n+1}}\enspace .
\end{align*}
By substituting $C_1=\frac{M\sqrt{\bar{m}}(L_{max}dC_{\alpha}+\sqrt{d}C_{\beta})}{2^{n+1}}$ and $C_2=\frac{\sqrt{2}}{2}\cdot\frac{M\sqrt{\bar{m}}C_{\alpha}}{2^{n+1}}$ and using the parameters defined in Assumption~\ref{as: conditions on initial intervals and the number of bits}, it follows that the expression above is equal to
\begin{align*}
 = \kappa^{g+1}\bigg[a_1 +a_2\frac{C_{\alpha}}{2^{n+1}} + a_3\cdot\frac{C_{\beta}}{2^{n+1}}\bigg]\enspace .
\end{align*}
By inequality~(\ref{eq:first condition on n and quantization intervals}) in Assumption~\ref{as: conditions on initial intervals and the number of bits}, the term above is bounded by $\frac{C_{\alpha}}{2}\kappa^{g+1} $. Thus, inequality~(\ref{eq: x falls in the interval at g+1}) holds. In the following, we prove that inequality (\ref{eq: the gradient of f falls in the interval at g+1}) is true.
\begin{align*}
\|\nabla f^{g+1}_i-\bar{\nabla} f^{g+1}_{\beta,i}\|_{\infty}&=\|\nabla f^{g+1}_i-\hat{\nabla} f^{g}_{i}\|_{\infty}\\
&= \|\nabla f_i(\tilde{x}^{g+1}_{\mathcal{N}_i})-\nabla f_i(\tilde{x}^{g}_{\mathcal{N}_i}) + \beta^{g}_i\|_{\infty} \\
&\leq \|\nabla f_i(\tilde{x}^{g+1}_{\mathcal{N}_i})-\nabla f_i(\tilde{x}^{g}_{\mathcal{N}_i})\|_{\infty} + \|\beta^{g}_i\|_{\infty} \\
&\leq \|\nabla f_i(\tilde{x}^{g+1}_{\mathcal{N}_i})-\nabla f_i(\tilde{x}^{g}_{\mathcal{N}_i})\|_{2} + \|\beta^{g}_i\|_{2} \\
&\leq L_i\|\tilde{x}^{g+1}_{\mathcal{N}_i}-\tilde{x}^{g}_{\mathcal{N}_i}\| + \|\beta^{g}_i\| \\
&\leq L_i\| x^{g+1}_{\mathcal{N}_i}- x^{g}_{\mathcal{N}_i} \| + L_i\|\tilde{x}^{g+1}_{\mathcal{N}_i} - x^{g+1}_{\mathcal{N}_i}\| +L_i\|\tilde{x}^{g}_{\mathcal{N}_i} - x^{g}_{\mathcal{N}_i}\| + \|\beta^{g}_i\|
\end{align*}

Since $x^{g+1}_{\mathcal{N}_i}, x^{g}_{\mathcal{N}_i}\in\mathbb{C}_{\mathcal{N}_i}$, $\tilde{x}^{g+1}_{\mathcal{N}_i} = \mbox{Proj}_{\mathbb{C}_{\mathcal{N}_i}}(\hat{x}^{g+1}_{\mathcal{N}_i})$ and $\tilde{x}^{g}_{\mathcal{N}_i} = \mbox{Proj}_{\mathbb{C}_{\mathcal{N}_i}}(\hat{x}^{g}_{\mathcal{N}_i})$, Lemma~\ref{le:proj(v) is closer to mu than v, for mu in C} implies $\|\tilde{x}^{g+1}_{\mathcal{N}_i} - x^{g+1}_{\mathcal{N}_i}\|\leq \|\hat{x}^{g+1}_{\mathcal{N}_i} - x^{g+1}_{\mathcal{N}_i}\|$ and $\|\tilde{x}^{g}_{\mathcal{N}_i} - x^{g}_{\mathcal{N}_i}\|\leq \|\hat{x}^{g}_{\mathcal{N}_i} - x^{g}_{\mathcal{N}_i}\|$. Hence, the term above is upper-bounded by

\begin{align*}
&\leq L_i\| x^{g+1}_{\mathcal{N}_i}- x^{g}_{\mathcal{N}_i}\| +  L_i\|\hat{x}^{g+1}_{\mathcal{N}_i} - x^{g+1}_{\mathcal{N}_i}\| +L_i\|\hat{x}^{g}_{\mathcal{N}_i} - x^{g}_{\mathcal{N}_i}\| + \|\beta^{g}_i\|\\
&\leq L_i\|x^{g+1}_{\mathcal{N}_i}-x^{g}_{\mathcal{N}_i}\| + L_i\sum_{j\in\mathcal{N}_i}(\|\alpha^{g+1}_j\| + \|\alpha^{g}_j\|)+\|\beta^{g}_i\| \\
&\leq L_i\|x^{g+1}-x^{g}\| + L_i\sum_{j\in\mathcal{N}_i}(\|\alpha^{g+1}_j\| + \|\alpha^{g}_j\|) +\|\beta^{g}_i\| \\
&\leq L_{\max}(\|x^{g+1}-x^{\star}\| + \|x^{g}-x^{\star}\|) + L_{\max}\sum_{j\in\mathcal{N}_i}(\|\alpha^{g+1}_j\| + \|\alpha^{g}_j\|) +\|\beta^{g}_i\| \enspace .
\end{align*}
Again by the assumption of the induction, we know $\|x^{k}_i-\bar{x}^{k}_{\alpha,i}\|_{\infty}\leq \frac{l^{k}_{\alpha,i}}{2}$ and $\|\nabla f^{k}_i-\bar{\nabla} f^{k}_{\beta,i}\|_{\infty}\leq \frac{l^{k}_{\beta,i}}{2}$ for $0\leq k\leq g$. Then, Lemma~\ref{le:linear decreasing rate of ek and xk} implies that the term above is upper-bounded by
\begin{align*}
\leq & L_{\max}\kappa^{g+1}\left(\|x^{0}-x^{\star}\|+\frac{(C_1+\sqrt{2L}C_2)\kappa}{L(\kappa+\gamma-1)}\right) +L_{\max}\kappa^{g}\left(\|x^{0}-x^{\star}\|+\frac{(C_1+\sqrt{2L}C_2)\kappa}{L(\kappa+\gamma-1)}\right) \\
& +\frac{L_{\max}\sqrt{\bar{m}}\sum_{j\in\mathcal{N}_{i}} (l^{g+1}_{\alpha ,j}+l^{g}_{\alpha ,j})}{2^{n+1}} + \frac{\sqrt{d\bar{m}}l^{g}_{\beta,i}}{2^{n+1}}\\
\leq & L_{\max}\kappa^{g+1}\left(\|x^{0}-x^{\star}\|+\frac{(C_1+\sqrt{2L}C_2)\kappa}{L(\kappa+\gamma-1)}\right) +L_{\max}\kappa^{g}\left(\|x^{0}-x^{\star}\|+\frac{(C_1+\sqrt{2L}C_2)\kappa}{L(\kappa+\gamma-1)}\right)\\
& +\frac{L_{\max}\sqrt{d\bar{m}}C_{\alpha}(\kappa^{g+1}+\kappa^{g})}{2^{n+1}} + \frac{\sqrt{d\bar{m}}C_{\beta}\kappa^{g}}{2^{n+1}}\enspace .
\end{align*}
By substituting $C_1=\frac{M\sqrt{\bar{m}}(L_{max}dC_{\alpha}+\sqrt{d}C_{\beta})}{2^{n+1}}$ and $C_2=\frac{\sqrt{2}}{2}\cdot\frac{M\sqrt{\bar{m}}C_{\alpha}}{2^{n+1}}$ and using the parameters defined in Assumption~\ref{as: conditions on initial intervals and the number of bits}, it follows that the expression above is equal to
\begin{align*}
= \kappa^{g+1}\cdot\bigg[b_1 + b_2\cdot \frac{C_{\alpha}}{2^{n+1}} + b_3\cdot\frac{C_{\beta}}{2^{n+1}}\bigg]\enspace .
\end{align*}
By inequality~(\ref{eq:second condition on n and quantization intervals}) in Assumption~\ref{as: conditions on initial intervals and the number of bits}, the term above is bounded by $\frac{C_{\beta}}{2}\kappa^{g+1}=\frac{l^{g+1}_{\beta,i}}{2}$. Thus, inequality~(\ref{eq: the gradient of f falls in the interval at g+1}) holds.
\end{itemize}
We conclude that by the principle of induction, the values of $x^{k}_i$ and $\nabla f^{k}_i$ in Algorithm~\ref{al:PGM for distributed optimization with quantization} fall inside of the quantization intervals of $Q^{k}_{\alpha,i}$ and $Q^{k}_{\beta,i}$, i.e. $\|x^{k}_i-\bar{x}^{k}_{\alpha,i}\|_{\infty}\leq \frac{l^{k}_{\alpha,i}}{2}$ and $\|\nabla f^{k}_i-\bar{\nabla} f^{k}_{\beta,i}\|_{\infty}\leq \frac{l^{k}_{\beta,i}}{2}$ for all $k\geq 0$.
\end{IEEEproof}


\subsection{Proof of Lemma~\ref{le: conditions on n and quantization intervals for APGM}}\label{SEC: appendix proof of Lemma of conditions on n and quantization intervals for APGM}

\begin{IEEEproof}
The proof is similar to the proof of Lemma~\ref{le: conditions on n and quantization intervals}. The difference is that at each iteration the gradient $\nabla f^{k}_i$ is computed based on $\tilde{y}^{k}_{\mathcal{N}_{i}}$, which is a linear combination of $\tilde{x}^{k}_{\mathcal{N}_{i}}$ and $\tilde{x}^{k-1}_{\mathcal{N}_{i}}$. We therefore only show a brief proof for the second step, i.e. the inequality $\|\nabla f^{k}_i-\bar{\nabla} f^{k}_{\beta,i}\|_{\infty}\leq \frac{l^{k}_{\beta,i}}{2}$ for any $k\geq 0$ by induction.
\begin{itemize}
\item Base case: When $k = 0$, since $C_{\beta}$ is positive a number, $\tilde{x}^{-1}_{\mathcal{N}_i}$ and $x^{0}_{i}$ are initialized to zero and $\hat{\nabla} f^{-1}_i = \nabla f_i(\mbox{Proj}_{\mathbb{C}_{\mathcal{N}_i}} (0))$, it holds that $\|\nabla f^{0}_i-\bar{\nabla} f^{0}_{\beta,i}\|_{\infty}= \|\nabla f^{0}_i-\hat{\nabla} f^{-1}_i\|_{\infty} = \|\nabla f_i(\tilde{y}^{0}_{\mathcal{N}_i})-\nabla f_i(\mbox{Proj}_{\mathbb{C}_{\mathcal{N}_i}} (0))\| = 0 \leq \frac{l^{0}_{\beta,i}}{2}= \frac{C_{\beta}}{2}$.
\item Induction step: Let $g\geq 0$ be given and suppose that $\|x^{k}_i-\bar{x}^{k}_{\alpha,i}\|_{\infty}\leq \frac{l^{k}_{\alpha,i}}{2}$ and $\|\nabla f^{k}_i-\bar{\nabla} f^{k}_{\beta,i}\|_{\infty}\leq \frac{l^{k}_{\beta,i}}{2}$ for $0\leq k\leq g$. We will prove
\begin{equation}\label{eq: the gradient of f falls in the interval at g+1 for accelerated case}
\|\nabla f^{g+1}_i-\bar{\nabla} f^{g+1}_{\beta,i}\|_{\infty}\leq \frac{l^{g+1}_{\beta,i}}{2}\enspace .
\end{equation}
From the algorithm, we know
\begin{align*}
\|\nabla f^{g+1}_i-\bar{\nabla} f^{g+1}_{\beta,i}\|_{\infty}&=\|\nabla f^{g+1}_i-\hat{\nabla} f^{g}_{i}\|_{\infty}\\
&= \|\nabla f_i(\tilde{y}^{g+1}_{\mathcal{N}_i})-\nabla f_i(\tilde{y}^{g}_{\mathcal{N}_i}) + \beta^{g}_i\|_{\infty} \\
&\leq L_i\| y^{g+1}_{\mathcal{N}_i}- y^{g}_{\mathcal{N}_i}\| +  L_i\|\hat{y}^{g+1}_{\mathcal{N}_i} - y^{g+1}_{\mathcal{N}_i}\| +L_i\|\hat{y}^{g}_{\mathcal{N}_i} - y^{g}_{\mathcal{N}_i}\| + \|\beta^{g}_i\|\enspace .
\end{align*}
By substituting $\hat{y}^{g}_{\mathcal{N}_i} = \frac{2}{1+\sqrt{\gamma}}\hat{x}^{g}_{\mathcal{N}_i} - \frac{1-\sqrt{\gamma}}{1+\sqrt{\gamma}}\hat{x}^{g-1}_{\mathcal{N}_i}$, $y^{g}_{\mathcal{N}_i} = \frac{2}{1+\sqrt{\gamma}}x^{g}_{\mathcal{N}_i} - \frac{1-\sqrt{\gamma}}{1+\sqrt{\gamma}}x^{g-1}_{\mathcal{N}_i}$ and $L_{max} := \max_{1\leq i \leq M} L_i$, and using the fact that $\frac{2}{1+\sqrt{\gamma}}\leq 2$ and $\frac{1-\sqrt{\gamma}}{1+\sqrt{\gamma}}\leq 1$, the expression above is upper-bounded by
\begin{align*}
&\leq L_{\max}(2\|x^{g+1}-x^{\star}\| + 3\|x^{g}-x^{\star}\|+ \|x^{g-1}-x^{\star}\|) + L_{\max}\sum_{j\in\mathcal{N}_i}(2\|\alpha^{g+1}_j\| + 3\|\alpha^{g}_j\| + \|\alpha^{g-1}_j\|) +\|\beta^{g}_i\| \enspace .
\end{align*}
By the assumption of the induction and Lemma~\ref{le:linear decreasing rate of ek and xk for APGM}, we obtain that the above is upper-bounded by
\begin{align*}
\leq & L_{\max}(2\kappa^{g+1} + 3\kappa^{g}+ \kappa^{g-1})\left[\frac{2\sqrt{\Phi(x^{0})-\Phi(x^{\star})}}{\sqrt{\sigma_{\phi}}} + \frac{(2C_3+2\sqrt{2L}C_4+\sqrt{2\sigma_{\phi}}C_4)\kappa}{\sigma_{\phi}(\kappa-\sqrt{1-\sqrt{\gamma}})\cdot \sqrt{1-\sqrt{\gamma}}}\right]\\
& +\frac{L_{\max}\sqrt{\bar{m}}d(2l^{g+1}_{\alpha ,j}+3l^{g}_{\alpha ,j} + l^{g-1}_{\alpha ,j})}{2^{n+1}} + \frac{\sqrt{d\bar{m}}l^{g}_{\beta,i}}{2^{n+1}}\enspace .
\end{align*}
By substituting $C_3=\frac{M\sqrt{\bar{m}}(3L_{max}dC_{\alpha}+\kappa \sqrt{d}C_{\beta})}{\kappa\cdot 2^{n+1}}$ and $C_4=\frac{\sqrt{2}}{2}\cdot\frac{M\sqrt{\bar{m}}C_{\alpha}}{2^{n+1}}$ and using the parameters defined in Assumption~\ref{as: conditions on initial intervals and the number of bits for the accelerated algorithm}, the expression becomes
\begin{align*}
= & \kappa^{g+1}\cdot\bigg[b_4 + b_5\cdot \frac{C_{\alpha}}{2^{n+1}} + b_6\cdot\frac{C_{\beta}}{2^{n+1}}\bigg]\enspace .
\end{align*}
By inequality~(\ref{eq:second condition on n and quantization intervals for accelerated case}) in Assumption~\ref{as: conditions on initial intervals and the number of bits for the accelerated algorithm}, the term above is bounded by $\frac{C_{\beta}}{2}\kappa^{g+1}=\frac{l^{g+1}_{\beta,i}}{2}$. Thus, the inequality $\|\nabla f^{g+1}_i-\bar{\nabla} f^{g+1}_{\beta,i}\|_{\infty}\leq \frac{l^{g+1}_{\beta,i}}{2}$ holds. The proof of the induction step is complete. 
\end{itemize}
By the principle of induction, we conclude that the inequality$\|\nabla f^{k}_i-\bar{\nabla} f^{k}_{\beta,i}\|_{\infty}\leq \frac{l^{k}_{\beta,i}}{2}$ holds for any $k\geq 0$. 
\end{IEEEproof} 


%

%
%
%
%


%
%
%
%


\ifCLASSOPTIONcaptionsoff
  \newpage
\fi



%

\bibliographystyle{plain}        
\bibliography{PGM_quantization} 

%

%
%
%




\end{document}